\documentclass{csmadapter}

\usepackage[normalem]{ulem}
\usepackage[colorinlistoftodos]{todonotes}
\usepackage{amsmath}
\usepackage{subfigure,placeins,float}

\jvol{XX}
\jnum{XX}
\paper{X}
\jmonth{September}
\jname{XX}
\pubyear{2023}

\begin{document}

 % \title{Live Traffic Control Experiment via Automated Vehicles\stitle{Methodology and Implementation of the Largest mobile traffic control experiment to date}}
\title{Designing, simulating, and performing the 100-AV field test for the CIRCLES consortium\stitle{Methodology and Implementation of the Largest mobile traffic control experiment to date}}

    \author{MOSTAFA AMELI\textsuperscript{\textsection, *},
    SEAN MCQUADE\textsuperscript{\textdagger\textdagger},
    JONATHAN W.~LEE\textsuperscript{*},
    MATT BUNTING\textsuperscript{\textdaggerdbl},
    MATT NICE\textsuperscript{\textdaggerdbl},
    HAN
    WANG\textsuperscript{**},
    WILL BARBOUR\textsuperscript{\textdaggerdbl},
    RYAN WEIGHTMAN\textsuperscript{\textdagger\textdagger}
    CHRIS DENARO\textsuperscript{\textdagger\textdagger}
    RYAN DELORENZO\textsuperscript{\textdaggerdbl\textdaggerdbl},
    SHARON HORNSTEIN\textsuperscript{\textbardbl},
    JON F. DAVIS\textsuperscript{*},
    DAN TIMSIT\textsuperscript{\textdagger},
    RILEY WAGNER\textsuperscript{\textdaggerdbl},
    RITA XU\textsuperscript{*},
    MALAIKA MAHMOOD\textsuperscript{\textdaggerdbl\textdaggerdbl},
    MIKAIL MAHMOOD\textsuperscript{\textdaggerdbl\textdaggerdbl},
    {MARIA LAURA} {DELLE MONACHE}\textsuperscript{**},
    BENJAMIN SEIBOLD\textsuperscript{\textbardbl\textbardbl},
    DAN WORK\textsuperscript{\textdaggerdbl},
    JONATHAN SPRINKLE\textsuperscript{\textdaggerdbl},
    BENEDETTO PICCOLI\textsuperscript{\textdagger\textdagger, \textdaggerdbl\textdaggerdbl},
    ALEXANDRE M.~BAYEN\textsuperscript{*}}
    
\affil{
    \textsuperscript{\textsection}--Université Gustave Eiffel, COSYS-GRETTIA\\
    *--University of California, Berkeley, Department of Electrical Engineering and Computer Sciences\\
     **--University of California, Berkeley, Department of Civil and Environmental Engineering\\
    \textsuperscript{\textdagger\textdagger}--Rutgers University-Camden, Center for Computational and Integrated Biology\\
    \textsuperscript{\textdaggerdbl\textdaggerdbl}--Rutgers University-Camden, Department of Mathematical Sciences\\
    % \textsuperscript{\textdagger}--\'Ecole normale sup\'erieure Paris-Saclay, Paris-Saclay University\\
    \textsuperscript{\textdaggerdbl}--Vanderbilt University Civil and Environmental Engineering\\
    \textsuperscript{\textbardbl}--General Motors, Control and Learning Team\\
    %\textsuperscript{\P}--Queen's University\\
    \textsuperscript{\textbardbl\textbardbl}--Temple University, Department of Mathematics\\
    \textsuperscript{\textdagger}--\'Ecole Polytechnique
    }
\maketitle

\begin{figure*}[h!]
    \centering    \includegraphics[width=\textwidth]{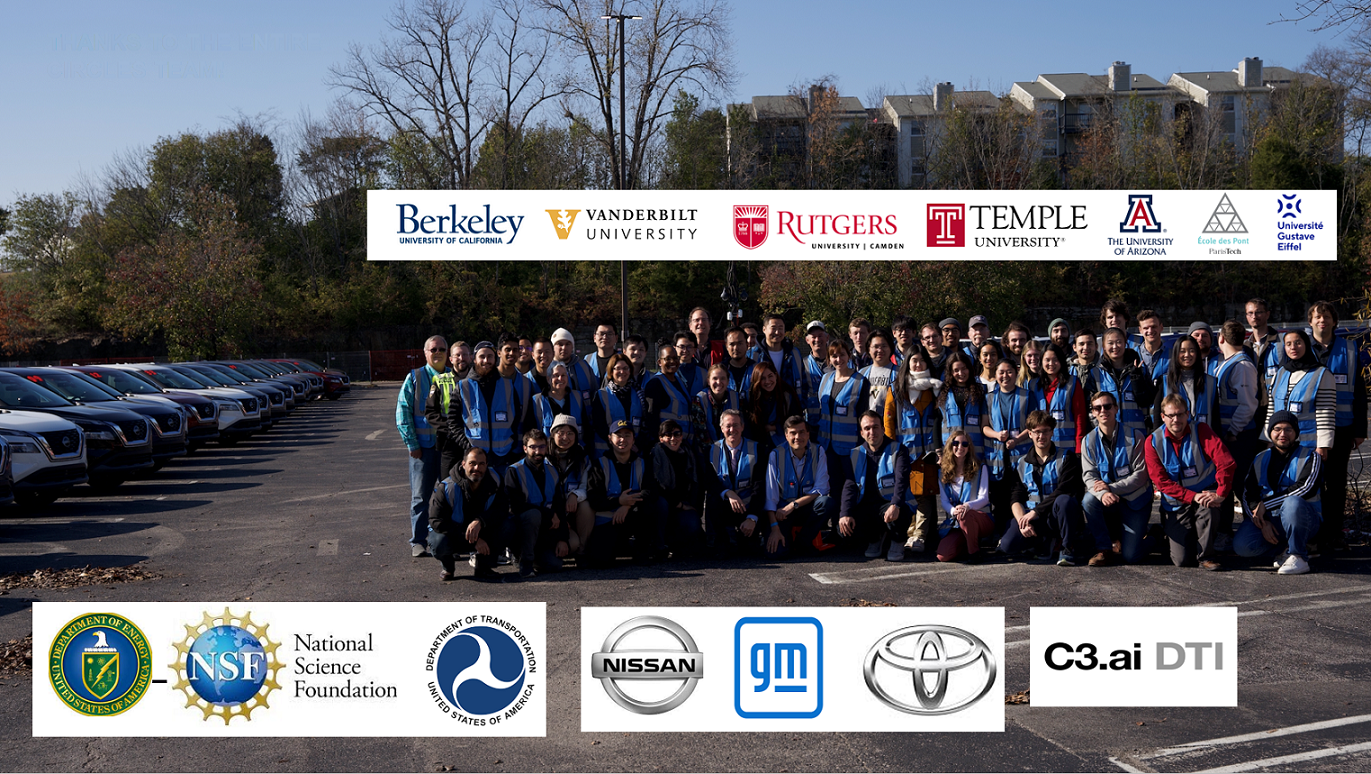}
    \caption{The CIRCLES Consortium at the experiment headquarters with partners listed.}
    \label{fig:CIRCLESteam}
\end{figure*}
% \dois{}{}

% The story of this paper will be as follows: we needed to set up an experiment on I-24. We needed to estimate certain aspects of traffic, such as queue length, vehicles on arterial roads, number of vehicles through traffic lights, etc. This necessitates a microsimulation to capture details. We needed to calibrate our microsim with Macro data, such as Inrix. The interperetation of this data requires PDE traffic estimation methods. The new introduction should be about the choices of experimental parameters (and maybe some microscopic models we could use).

% HACK: putting abstract on next page
\newpage
\begin{summary}
\summaryinitial{P}revious controlled experiments on single-lane ring roads \cite{wu2015cellpath} have shown that a single \textit{partially autonomous vehicle} (AV) can effectively mitigate traffic waves. This naturally prompts the question of how these findings can be generalized to field operational, high-density traffic conditions. To address this question, the \textit{Congestion Impacts Reduction via CAV-in-the-loop Lagrangian Energy Smoothing} (CIRCLES) Consortium conducted \textit{MegaVanderTest} (MVT), a live traffic control experiment involving 100 vehicles near Nashville, TN, USA. The purpose was to implement various controllers to smooth stop-and-go traffic waves.

This article is a tutorial for developing analytical and simulation-based tools essential for designing and executing a live traffic control experiment like the MVT. It presents an overview of the proposed roadmap and various procedures used in designing, monitoring, and conducting the MVT, which is the largest mobile traffic control experiment at the time. The design process is aimed at evaluating the impact of the CIRCLES AVs on surrounding traffic. The article discusses the agent-based traffic simulation framework created for this evaluation. A novel methodological framework is introduced to calibrate this microsimulation, aiming to accurately capture traffic dynamics and assess the impact of adding 100 vehicles to existing traffic. The calibration model's effectiveness is verified using data from a six-mile section of Nashville's I-24 highway. The results indicate that the proposed model establishes an effective feedback loop between the optimizer and the simulator, thereby calibrating flow and speed with different spatiotemporal characteristics to minimize the error between simulated and real-world data. Finally, We simulate AVs in multiple scenarios to assess their effect on traffic congestion. This evaluation validates the AV routes, thereby contributing to the execution of a safe and successful live traffic control experiment via AVs.

% Our methods capitalized on the advanced I-24 MOTION camera system \cite{gloudemans2020interstate}, which provided high-resolution insights into realtime traffic conditions. 

% based on two main data sets (commonly) which are used, including Inrix speed data, averaged across all lanes and flow data for highway segments that comprise the testbed (and adjacent arterial roads). 
\end{summary}

\chapterinitial{T}raffic congestion, particularly during rush hours, is a pervasive issue that results in substantial direct and indirect costs while also affecting fuel efficiency and road safety. In this context, vehicular traffic control can play a significant role in alleviating congestion. Although traffic models have been in development since the middle of the 20th century, focusing on macroscopic and microscopic dimensions, experimental efforts to understand traffic flow began with Bruce D. Greenshields in 1933 \cite{kuhne2008foundations}. He utilized a camera mounted on a mobile platform to precisely record the movements of a single vehicle. This groundbreaking experiment laid the foundation for subsequent studies aimed at evaluating the effectiveness of traffic models.

While there have been promising results in both theoretical and applied settings with controlled environments, live-highway experiments remain relatively rare. Over the past two decades, the focus of these experiments has shifted from merely measuring and understanding vehicle flow to actively attempting to influence it. This change has been facilitated by \textit{partially automated vehicles} (AVs); these AVs have a driver at all times and are limited to longitudinal control. Researchers can leverage AVs to test the implementation of new ideas, technologies, and algorithms designed to improve traffic conditions  \cite{DI2021103008, fu2023cooperative, lichtle2022deploying, wu2017flow, STERN2018205, STERN2019351, vinitsky2018benchmarks, wu2017emergent, Kardous2022multi, Hayat2023theory, Hayat2022holistic, lee2021integrated, wu2021flow, yan2022unified, bhadani2022strym, giammarino2020traffic}.

A practical understanding of the formation of traffic jams in the absence of bottlenecks was achieved by \cite{Sugiyama2008traffic}, conducting an experiment on a ring road with a circumference of 230 meters and 22 vehicles. A camera situated at the center of the circle recorded the movements of the vehicles. The study found that small variations in the speed of leading vehicles were amplified by the vehicles following them, thereby propagating a wave. These findings are consistent with theoretical microscopic models.

In 2010, the Mobile Century field experiment \cite{DanWork2010gps} used GPS technology in order to track mobile phones inside vehicles traveling on California's I-880 freeway. This experiment served as a proof of concept for the use of GPS technologies in real-world traffic studies. It demonstrated that key traffic characteristics, such as speed, could be effectively captured \cite{HERRERA2010568}.

Building on these foundational experiments, efforts have emerged to actively control traffic, typically in similar settings such as single-lane circular roads. The experiment detailed in \cite{stern2018dissipation} serves as a proof of concept for smoothing traffic waves to reduce energy costs for all vehicles on the road. Using a circular track with 20 vehicles, it was observed that traffic waves naturally form when vehicles are solely human-controlled, as previously demonstrated in \cite{Sugiyama2008traffic}. However, these waves can be effectively dissipated by introducing a small fraction of controlled vehicles into the mix. Remarkably, the experiment showed significant wave dampening with just one vehicle equipped with a control algorithm among 20 human-driven counterparts.

This development raises questions about the feasibility of replicating these controlled experiments in actual highway settings. Specifically, it prompts inquiries regarding the required penetration rate of AVs relative to \textit{human-driven vehicles} (HDVs), as well as the concept of vehicle platooning, where vehicles are closely grouped together. The current article focuses on the design and execution of a large-scale live traffic control experiment aiming to replicate the findings of the controlled circular road experiment and quantify specific outcomes, such as energy savings.

\begin{figure}[!t]
    \centering
    \subfigure[Mapping data \textcopyright Google maps.]
    {
        \includegraphics[width=3.0in]{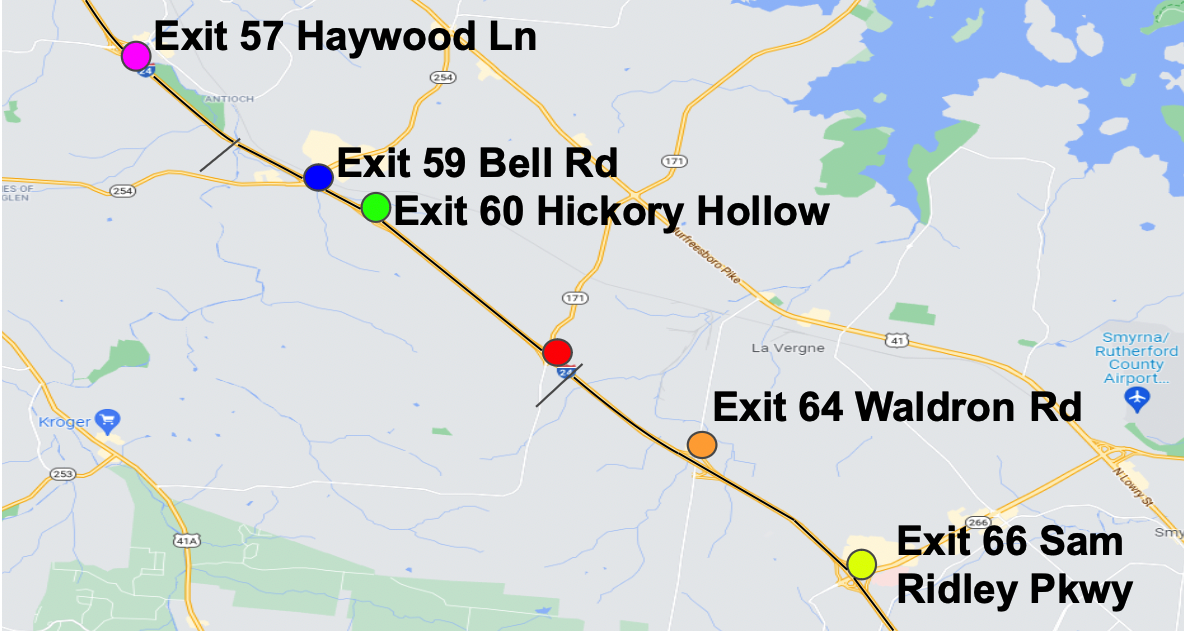}
        \label{fig:I-24_GM}
    }
    \subfigure[SUMO Microsimulation Network.]
    {
        \includegraphics[width=3.0in]{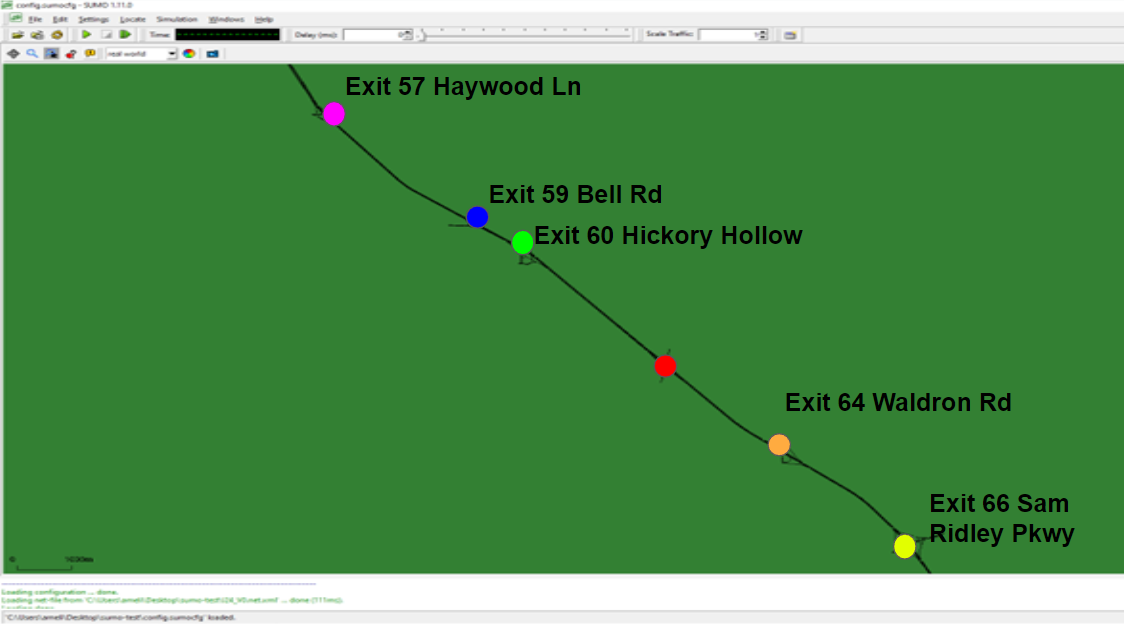}
        \label{sumo:network}
    }
    \subfigure[Sample of intersections of I-24 SUMO network verified with online mapping data.]
    {
        \includegraphics[scale = 0.2]{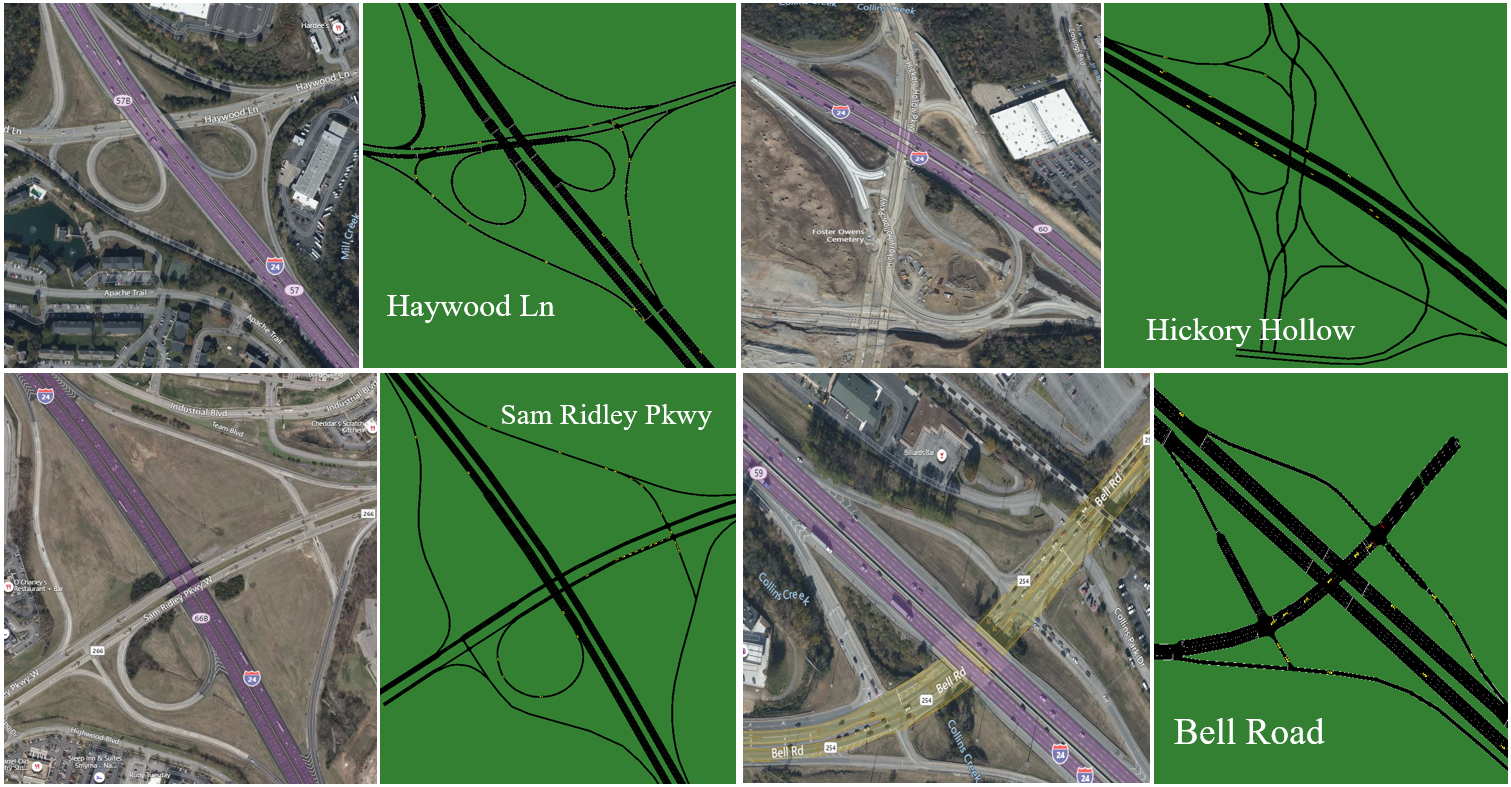}
        \label{sumo:intersections}
    }
    \caption{I-24 road network: The MVT experiment testbed shown with \textcopyright Google maps and SUMO .}
    \label{fig:I-24}
\end{figure}

This live experiment, referred to as the \textit{MegaVanderTest} (MVT), was conducted on Interstate 24 West towards Nashville, TN, depicted in Figure \ref{fig:I-24_GM}. Designed over a three-year period, MVT aimed to alleviate stop-and-go traffic waves using AVs. This article elaborates on the comprehensive planning that included experiment design, simulation tests, and driver training. Leveraging state-of-the-art design validation tools, monitoring systems, and control algorithms, 100 AVs were deployed to navigate and influence traffic flow. The flexible design allowed for real-time adaptability to both foreseen and unforeseen obstacles, such as needing to hire more drivers as we approached the test and off-loading 100 dashcams worth of data as the storage became full. Preliminary analyses suggest a measurable impact on highway traffic, contributing to safer and more efficient road usage.

%%%%%%%%%%

\section{The ``MegaVanderTest'' Experiment}
Building on insights gained from ring road experiments \cite{stern2018dissipation} and a small 4-AV test in 2021 \cite{lichtle2022deploying}, the main 100 AV experiment run by the CIRCLES Consortium, named the MVT, aims to intervene in the bulk traffic flow on \textit{Interstate 24} (I-24) West towards Nashville, TN. Unlike traditional adaptive cruise control systems, which are string-unstable and exacerbate traffic waves \cite{gunter2020commercially}, MVT employs a fleet of 100 AVs enhanced with dynamically changing cruise control to implement real-time, adaptive interventions. This technology is designed to react to immediate leader vehicles (the vehicle in front), though differently than typical Adaptive Cruise Control (ACC), but also it is able to utilize downstream traffic data to anticipate major slow downs. Key metrics we aim to measure include changes in energy consumption, traffic speed variability, and the frequency of braking events. The I-24 MOTION system~\cite{gloudemans202324,gloudemans2020interstate} serves as the monitoring infrastructure, capturing high-resolution data for rigorous post-analysis of the effect on the surrounding traffic. The instrument captures approximately 230 million vehicle-miles of travel annually, and experiences regular recurring congestion. The MOTION system~\cite{gloudemans2023so,gloudemans2021vehicle,gloudemans2023interstate} consists of a computer vision pipeline, a trajectory post processing pipeline~\cite{wang2022automatic,wang2023onlinemcf}, and a visualization tool~\cite{10.1145/3576914.3587710}.

\begin{figure}[hb!]
    \centering
    \includegraphics[width=\columnwidth]{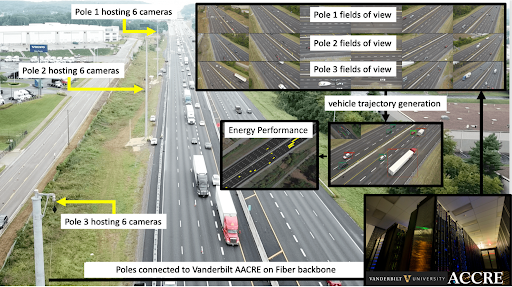}
    \caption{The I-24 MOTION system~\cite{gloudemans202324,gloudemans2020interstate} comprises 276 cameras mounted on 40 poles ranging from 110 ft to 135 ft above the freeway along a 4.2 mile stretch of Interstate 24, southeast of Nashville, Tennessee.}
    \label{fig:I24MOTION}
\end{figure}

The mid-November morning traffic on I-24 highway in 2022 seemed indistinguishable from typical rush hours, densely packed with vehicles. The AVs were designed to prevent abrupt "jack-rabbit" starts and stops often observed during congested traffic periods \cite{JangReinforcementCSM}. These AVs appear to influence the driving behavior of the following cars, potentially mitigating intense traffic fluctuations and promoting a more consistent flow. Initial observations, based on data from the first few days of the experiment week, suggest that even a minor presence of AVs might have the potential to moderate stop-and-go patterns \cite{LeeMegacontrollerCSM}, which could lead to better energy efficiency for the majority of vehicles \cite{KhoudariEnergyCSM}. This potential benefit might not just be restricted to energy; it could also result in reduced emissions and possibly improve road safety \cite{HayatMicrocontrollerCSM}. It's worth noting that past studies have pointed out a 36\% surge in crash injury risk during high-traffic periods in addiction to traffic wave alleviation \cite{8569615, fu2023cooperative, kheterpal2018flow}.

% However, upon closer scrutiny, facilitated by our data analytics, a fraction of the vehicles exhibited more stable driving patterns, eschewing the "jack-rabbit" starts and stops that plague congested traffic \cite{JangReinforcement}. Our control vehicles, the AVs, influenced the driving patterns of the vehicles following them, effectively damping traffic waves and generating a more constant flow. Preliminary analysis of the data from the MVT suggests that even a minimal introduction of AVs can smooth stop-and-go waves to yield significant energy gains for the bulk traffic \cite{LeeMegacontroller}. This smoothing effect also extends beyond energy efficiency;  it lowers emissions and positively influences road safety by reducing the risk of accidents. This is particularly significant given that previous research indicates a 36\% increase in crash injury risk during peak traffic hours \cite{HayatMicrocontroller}. 

\section{MVT Experiment design}
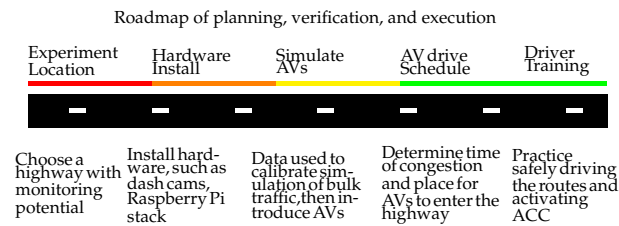
\begin{figure}[!b]
\scalebox{0.55}{
\begin{tikzpicture}
     % Road surface (black)
    %[[text width=5mm]]
    \tiny
        \node (title) at (6.7,5) {\Large Roadmap of planning, verification, and execution};
        \node[text width=2cm] (experiment) at (1,4) {\Large Experiment Location};
        \node[text width=2cm] (install) at (4,4) {\Large  Hardware Install};
        \node[text width=20mm] (design) at (7,4) {\Large Simulate AVs};
        \node[text width=20mm] (routes) at (10,4) {\Large AV drive Schedule};
        \node[text width=2cm] (training) at (13,4) {\Large Driver Training};

        \node[align=left][text width=26mm] (exp_desc) at (1,1.0) {\Large Choose a highway with monitoring potential};
        \node[align=left][text width=28mm] (install_desc) at (3.8,1.0) {\Large Install hardware, such as dash cams, Raspberry Pi stack};
        \node[align=left][text width=28mm] (design_desc) at (6.8,1.0) {\Large Data used to calibrate simulation of bulk traffic,then introduce AVs};
        \node[align=left][text width=29mm] (routes_desc) at (10,1.0) {\Large Determine time of congestion and place for AVs to enter the highway};
        \node[align=left][text width=26mm] (training_desc) at (13,1.0) {\Large Practice safely driving the routes and activating ACC};
    
    %colored underlines
      \fill[color=red] (0,3.4) rectangle (3,3.5);
      \fill[color=orange] (3,3.4) rectangle (6,3.5);
      \fill[color=yellow] (6,3.4) rectangle (9,3.5);
      \fill[color=green] (9,3.4) rectangle (14,3.5);
      %road
      \fill[color=black] (0,2.4) rectangle (14,3.2);
      % Lane markings
      \foreach \y in {2.75}
        \foreach \x in {1,3,5,7,9,11,13}
          \fill[color=white] (\x,\y) rectangle ++(0.4,.1);
\end{tikzpicture}
}
\caption{Road map showing the planning and execution sequence of project milestones.} 
\label{fig:roadmap}   
\end{figure}

%Being the largest mobile traffic control experiment of its time, 
The planning and preparation for the MVT were extensive. Devised and spearheaded by the CIRCLES Consortium, steps were rigorously executed to ensure participant safety and to maximize the penetration rate of AVs within the I-24 MOTION testbed. The primary objective was to assess the impact of the AVs on general traffic behavior. Coordinating a project of this scale, which involved an interdisciplinary team from multiple universities, auto manufacturers engineers, along with others, required meticulous planning, installation, and simulation, as illustrated by the project road map (Figure \ref{fig:roadmap}). 

This article focuses on the methodological framework used to design and conduct the MVT experiment. We outline the comprehensive three-year planning process and preparation, which includes the selection of experimental dates and times, route planning, hardware installation, and developing and simulating control algorithms. Drivers were trained in order to operate the control vehicles both safely and efficiently, and dynamic agent-based simulations were utilized to assess various scenarios and refine the experimental design (see Figure \ref{fig:I-24}). 

The first two steps in Figure \ref{fig:roadmap} are detailed through sidebars: step 1 is elaborated in "MegaVanderTest testbed location," and step 2 in "Integrated hardware installation overview." ``Daily workflow and AV release schedule,'' step 3 in sections ``Methodological Framework to Generate the Background Traffic to Conclusion'' step 4 in ``MegaVanderTest testbed location'' and ``Optimized Deployment Schedule'' step 5 in ``Driver safety procedures and training.'' The following section details step 5 of the road map. Here the creation and calibration of the agent-based simulation framework to represent actual traffic is presented in detail. This was used to design and evaluate the AV routes and driving schedule for the MVT experiment, as well as important details like the clustering of AVs, additional queue length, congestion on arterial roads, and the potential for queues to overflow off-ramps. The sidebar, ``Optimized vehicle deployment and online monitoring system'' details the live monitoring system needed to conduct the MVT experiment. This allows us to know the status of the AVs at all times, keeping all AVs and their drivers accounted for.

% \begin{figure}[!b]
% \includegraphics[scale=0.11, trim={1.2cm 5cm 3.2cm 0.6}, clip]{figures/MyVersionRoadMapV22.png}
% \caption{Road map showing the planning and execution sequence of project milestones. {\footnotesize(image template: www.freepik.com)}}
% \label{fig:roadmap}
% \end{figure}

Detailed planning covered various aspects, including the design of AV routes, their entry points onto I-24, and measuring the impact on congestion near traffic lights. We also evaluated the potential impact of 100 AVs driving in repetitive loops through the testbed, taking into account factors such as queue lengths at on/off ramps, traffic light-induced congestion, and the clustering of AVs.

\begin{sidebar}{MegaVanderTest testbed location}
\noindent
by Sean T. McQuade and Mostafa Ameli
\setcounter{sequation}{0}
\renewcommand{\thesequation}{S\arabic{sequation}}
\setcounter{stable}{0}
\renewcommand{\thestable}{S\arabic{stable}}
\setcounter{sfigure}{0}
\renewcommand{\thesfigure}{S\arabic{sfigure}}

\noindent
\sdbarfig{\label{fig:DawnLot}\includegraphics[width=19.0pc]{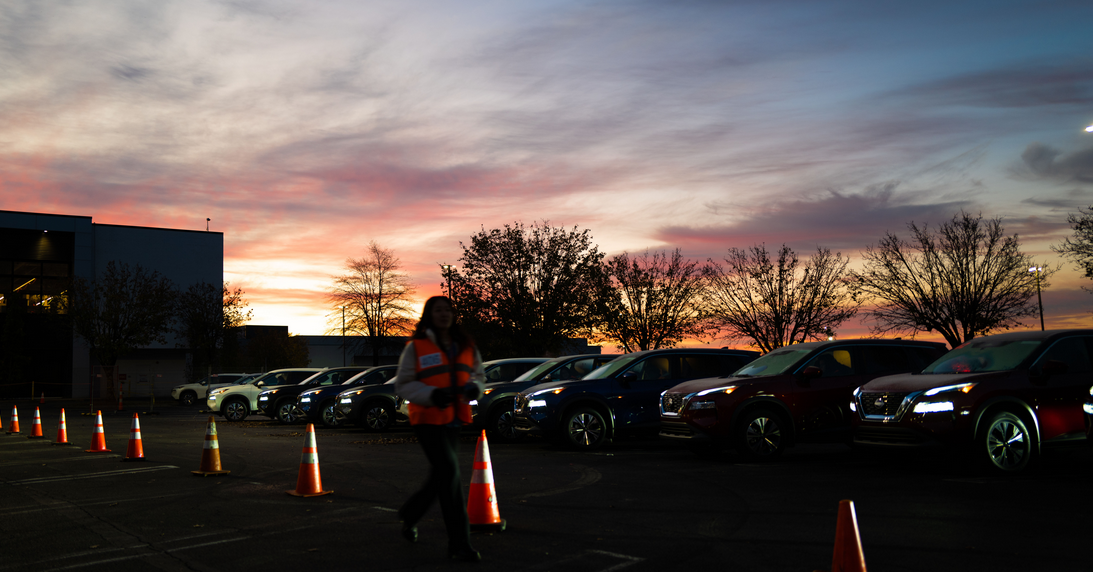}}{Parking space of 100 AVs during the MegaVanderTest experiment.}

\sdbarinitial{T}\emph{he} Interstate 24 (I-24) MOTION testbed~\cite{gloudemans2020interstate,gloudemans2021vehicle} is located Southeast of Nashville. The locatio was chosen to capture rush hour traffic dynamics including recurring stop and go driving. This is also ideal to test the ability to smooth traffic waves in congestion, and it became the testbed for the MVT experiment. Historical data was used to determine when the peak congestion would occur on the testbed. The CIRCLES Consortium decided that these conditions lead to an abundance of stop-and-go traffic waves, and this gives us the highest chance for the algorithms to smooth traffic.

To positively impact the traffic flow, there must be sufficient partially \textit{autonomous vehicles} (AVs) relative to background traffic on the road \cite{WangSpeedplannerCSM}. The metric quantifying this is called the penetration rate, which is defined as the proportion of AVs of the total traffic (further detailed in section "Penetration rate estimations"). We estimated the penetration rate we could achieve with various strategies using an official report from the \textit{Tennessee Department of Transportation} (TDOT). This report, titled ``I-24 Ramp Metering Study (From I-840 to I-40) Rutherford and Davidson Counties Traffic Operations Report
July 2021,'' provided counts of vehicles on parts of I-24 and the on/off ramps. This data served as the basis for estimating the penetration rate and the total vehicles through traffic lights.

The initial strategy to maximize the penetration rate of the 100 AVs was to release them to I-24 just before rush hour and have the vehicles repeatedly drive from exit 66 to exit 57 westbound(WB), turn around at this exit, and drive eastbound(EB) returning to exit 66. This route requires that the AVs would drive through four traffic lights (two at each end of the route to turn around; the exit 57 traffic lights are labeled in Figure \ref{fig:TDOT_I24_volume} as black dots 7, and 8). Two key considerations were: (1) would the extra vehicles overflow the traffic light and result in a long queue forming on the highway, and (2) would the AVs cluster at the lights since they would be a large proportion of the vehicles that make two lefts consecutively to return to I-24 EB from I-24 WB (likely close to 100\%).

It was found that all 100 AVs driving this route would add approximately 176 control vehicles per hour to each traffic light (the loop would take approximately 34 minutes). Comparing this to traffic light 7 from Figure \ref{fig:TDOT_I24_volume} reportedly has 355 vehicles making the left turn onto Haywood lane from the I-24 off ramp during 6:30-7:30. The team decided that this would likely produce a large queue and disrupt traffic. To reduce the adverse impact on traffic light queuing, we decided to have two separate routes. The two routes would overlap on the road where the highest penetration is needed, but they have different end points, distributing the AVs among more traffic lights. The two routes, called the ``orange'' and ``yellow'' routes (exit 64 to 57, exit 66 to 60 respectively), are shown in Figure \ref{fig:Routes}. Drivers were given QR codes, and their phones were plugged into the center console of the AVs to display the route on screen.

\sdbarfig{\label{fig:Routes}\includegraphics[width=19.0pc]{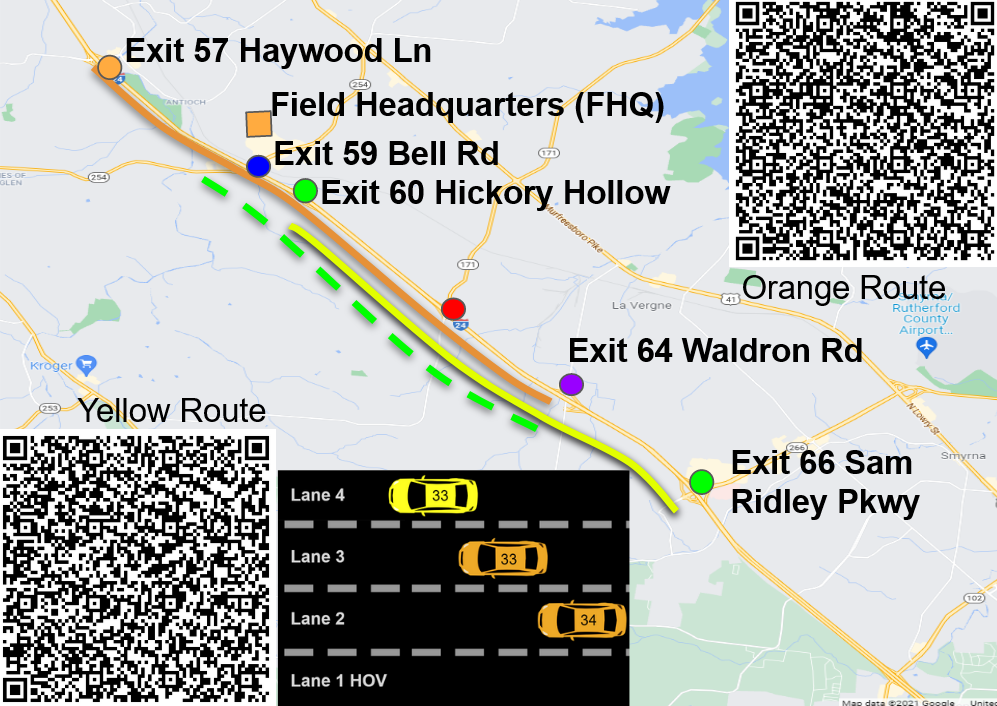}}{Two routes (orange and yellow) and I-24 MOTION~\cite{gloudemans202324,gloudemans2020interstate} coverage (green dashed). AVs only occupied lanes 2, 3, and 4. The QR codes that were provided to drivers are shown.}
\end{sidebar}

To further understand these complex variables, we employed the \textit{Simulation of Urban MObility} (SUMO) environment \cite{SUMO2018}. While agent-based models allow for detailed representations of individual vehicle behavior, the calibration of these models presents challenges due to the limitations of available data sets. For the MVT simulation framework, calibration relied on flow data from the \textit{Tennessee Department of Transportation} (TDOT) and speed data from INRIX \cite{cookson2017inrix}. A primary challenge lies in the differing spatiotemporal characteristics between these two data sets. Subsequent sections will illustrate how the proposed framework addresses this challenge in order to accurately represent traffic dynamics.

\setcounter{figure}{4}
\begin{figure}[h!]
    \centering
    \includegraphics[scale=0.5, trim={0 5cm 20cm 0}, clip ]{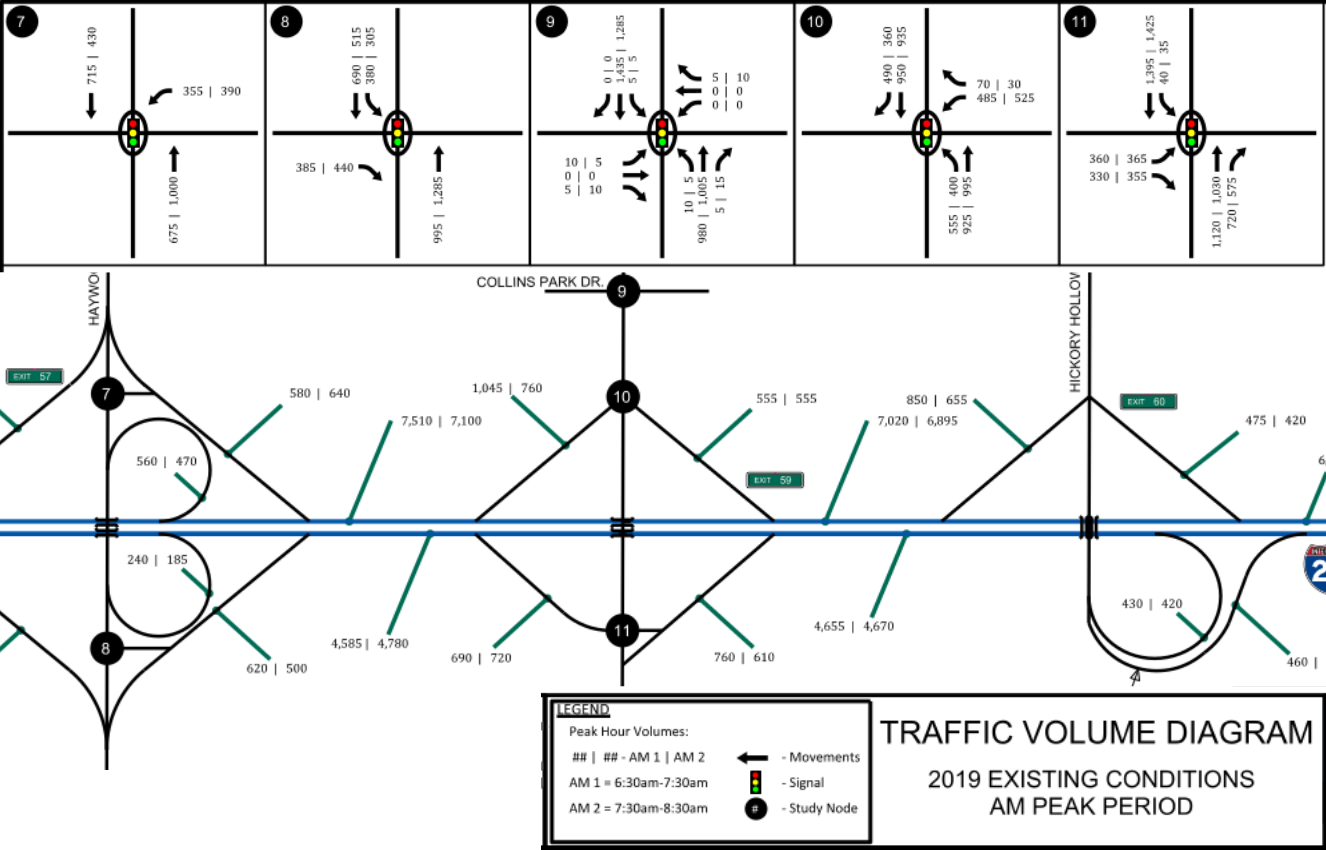}
    \caption{Throughput data for July 2021. 
    The pairs of numbers indicate how many vehicles drove through the section of road indicated by the green line during peak congestion. The number on the left shows the count of vehicles between 6:30am to 7:30am, and the number on the right shows the count between 7:30am and 8:30am.
    %These numbers indicate the amount of vehicles passing through on sections of I-24, ramps, and nearby traffic lights during peak congestion. The paired numbers indicate vehicle counts from two peak hours, 6:30-7:30am $|$ 7:30-8:30am.
    }
    \label{fig:TDOT_I24_volume}
\end{figure}

\begin{sidebar}{Integrated hardware installation overview}
\noindent
by Matt Bunting
\setcounter{sfigure}{2}
\renewcommand{\thesfigure}{S\arabic{sfigure}}

    \sdbarinitial{V}\emph{ehicles} needed to be outfitted with various components to enable control and collect data.  When outfitting the very first vehicle, it took one person about 5 hours to perform the installation task.  Scaling this to 100 vehicles results in 20 days of continuous effort.  Certainly increasing personnel can reduce the installation time, however training takes time and training too many tasks can lead to missed steps.

    Various hardware components arrived on different days prior to the experiment, meaning hardware installation had to be dynamically adapted to  resources currently available.  One of the final components, a libpanda-supported custom PCB named the ``mattHat,'' was obtained prior to the first experiment \cite{bunting2021libpanda}.  Teams were created to perform different tasks based on the hardware availability and task complication.  Some tasks required multiple people, like the removal of the center console due to its weight and position.  Training for the center console removal can be seen in figure~\ref{fig:install-training}.  Following the factory service manual also required procedures like disconnecting the battery when working around the airbag or when disconnecting electrical modules to install a custom wire harness.  

    \sdbarfig{\label{fig:install-training}\includegraphics[width=19.0pc]{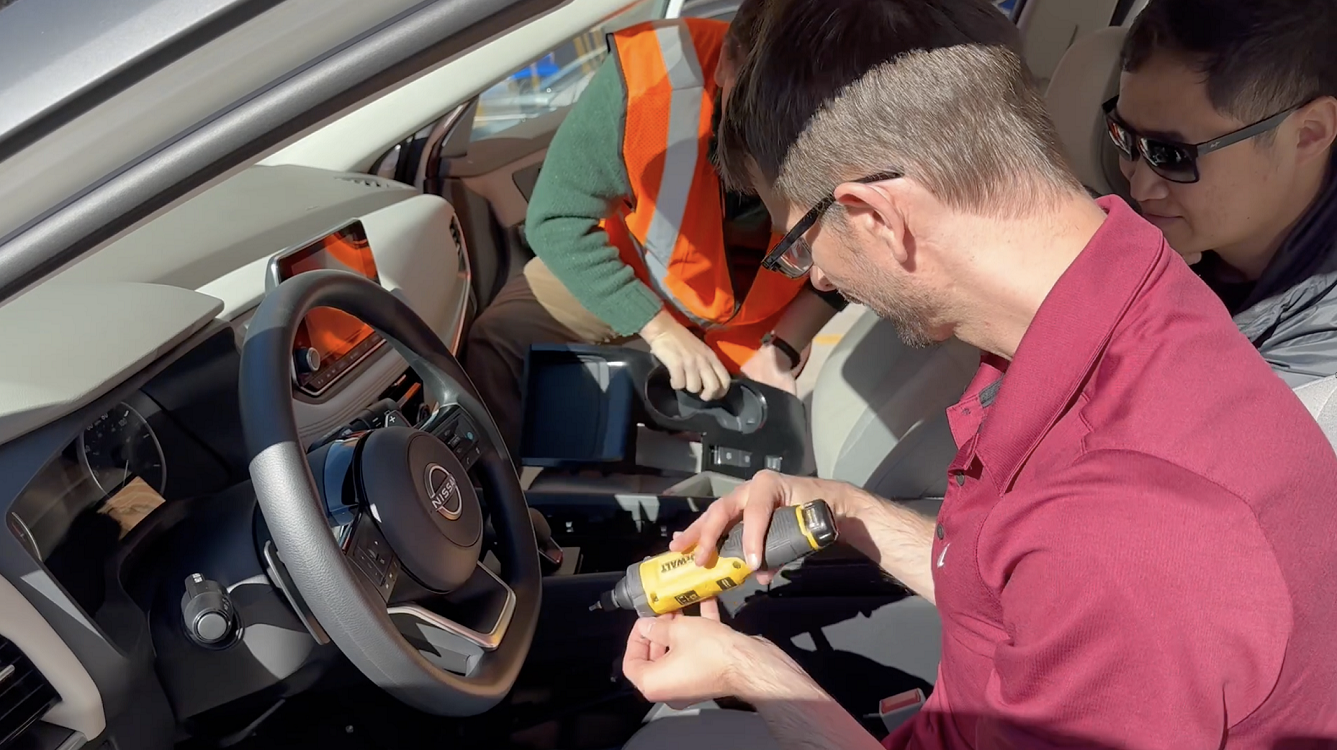}}{Training on removing the center console to gain access to the vehicle's wire harness so that custom hardware can be attached. Sean McQuade (center) is demonstrating for Tianya Zhang (right)}

    The set of major tasks to install the hardware were as follows:
    \begin{itemize}
        \item Dashcams - Wires needed to be routed out of view of the driver, meaning along the A-pillar which contained a passenger airbag. The battery had to be disconnected according to the factory service manual.
        \item Data Recording - This required tapping into the vehicle's CAN bus harness near an electrical module hidden under the center console.  Disconnection of the electrical module also required a battery disconnect.
        \item ACC wire for control - A particular wire harness connector was located in the driver foot well quarter panel, requiring removal of the hood latch.  A solid core wire (wire intended for home door bells) was then inserted into the connector so the other end could attach to the ``mattHat.''
        \item Generic cable routing - Various cables had to be routed in a manner that would not cause safety issues, like obstructing the driver's view or mechanically compromising the vehicle controls, like the pedals.
        \item Raspberry Pi Stack - the main hardware component involved a Raspberry Pi along with HAT (Hardware Attached on Top) modules that could be pre-assembled in the headquarters and brought to each vehicle in a marked tote.  This configuration process also included flashing software.
        \item Cradlepoint MiFi - These devices were on a generous loan and needed to be returned in factory-perfect condition, meaning that they were given special careful handling when being transferred to and out of the vehicles.
    \end{itemize}

    While installation was a time consuming task that took weeks to accomplish, a secondary daunting task was the removal process of all hardware to return the vehicles to a factory state for their return.  Fortunately, many of the tasks for hardware removal followed the same procedure for installation with the benefit that wires were getting removed instead of neatly run and that there were no holdups on obtaining supplies.  Teams were already experienced in doing such a task, however we were motivated to return the vehicles as quickly as possible.

    % Deconstruction was planned to be performed in a mere 2 days, versus the weeks to install.  This means that there was careful consideration in that tasks that could be performed on the first day so that the second day could be done efficiently.  The major efficiency component to plan around was the disconnection of the vehicle battery. vehicles needed to be locked for the night, requiring power to be connected for the electronic locks.  To prevent the need to reconnect the battery at the end of day 1 only to disconnect it again on day 2, tasks were given that led up to the point just prior to the battery disconnection.  Similarly, the hood latch had to be removed for the ACC wire, which was needed to grant access to disconnect the battery, so this was saved for day 2 as well.  

    \sdbarfig{\includegraphics[width=12.0pc]{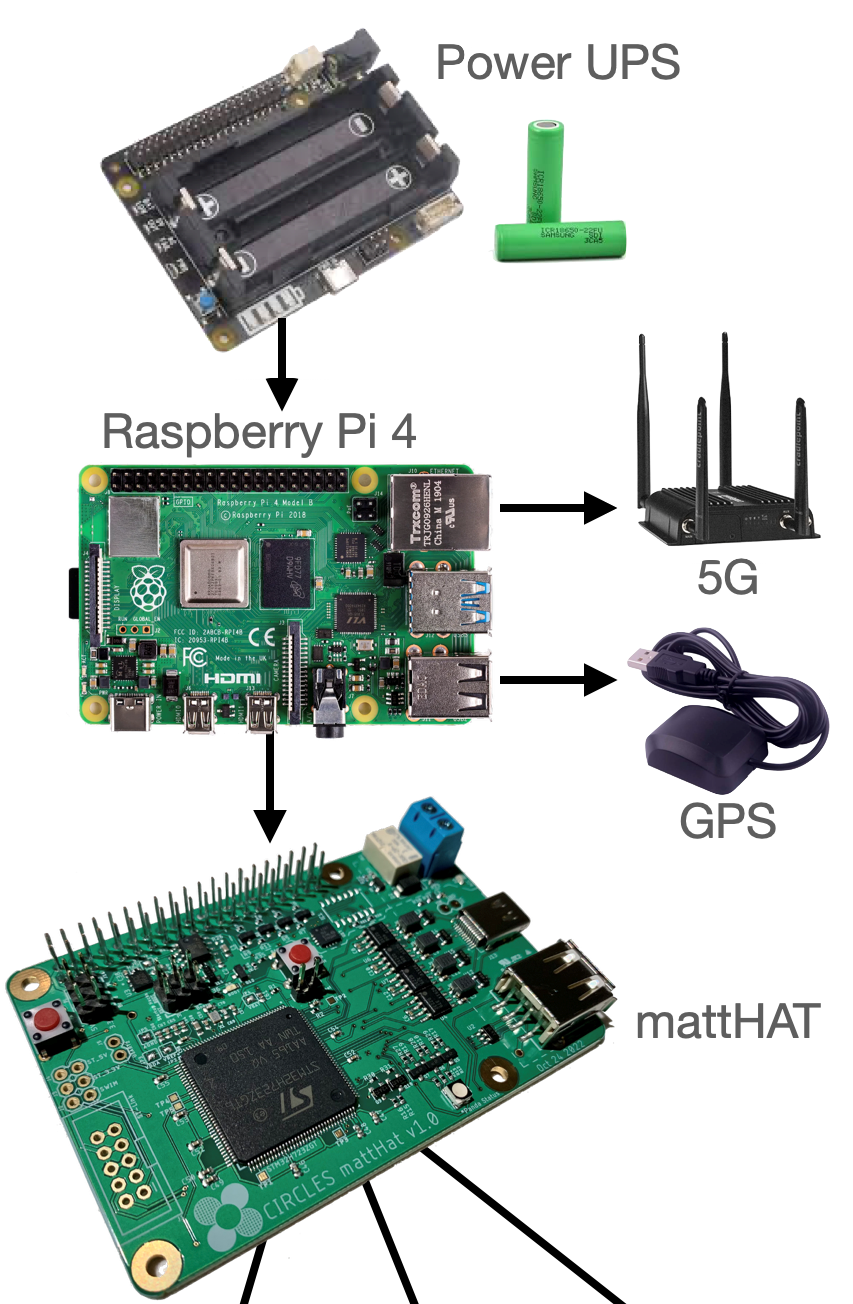}}{The hardware installed in each vehicle along with dash cams. \label{sfig_hardware_stack}}

    To speed up the hardware removal process even further along with handling new volunteers, tasks were highly compartmentalized.  A leader would task each volunteer with a minimal task to be performed on all vehicles.  This created an assembly-line process where individuals could be become highly skilled with a basic task rather than needing to remember every step for the whole installation.  This process allowed the team to succeed in the hardware removal of all vehicles in only 2 days, and all vehicles were returned shortly after.
\end{sidebar}

\begin{sidebar}{Optimized vehicle deployment and online monitoring system}
\noindent
by Matt Bunting
\renewcommand{\thesfigure}{S\arabic{sfigure}}

    \sdbarinitial{T}\emph{he} hardware solutions for installing custom controllers involved considerations of scale at all stages, from design to deployment. From the perspective of system design, hardware had to be low cost and easy to install.  Low cost hardware that is rapidly installed can cause concerns about functional yield.  Therefore from the perspective of deployment, verification systems need to check that all hardware is functional before sending a vehicle into an experiment cycle.  Such systems also need to radio their status to the headquarters for clear, organized checking.  Lastly, given the agile development nature of a research experiment, automatic software updates were implemented for bug fixes or changing control parameters in subsequent experiment days.

    Each vehicle was outfitted with off-the-shelf components including items like mobile internet MiFi hotspots, Raspberry Pis, Raspberry Pi Battery Backup UPSs, and power adapters and cables.  Custom components included a wire harness that connects to the vehicle's infrastructure, and a custom circuit board for reading sensors and sending control messages.  If any of these individual devices failed or disconnected in a vehicle, then that vehicle would be unusable for the experiment.  This includes trivial problems like a cigarette lighter power adapter not being fully inserted; while trivial, it is a time consuming task to check on all 100 vehicles before each experiment.

\noindent
\sdbarfig{\includegraphics[width=19.0pc]{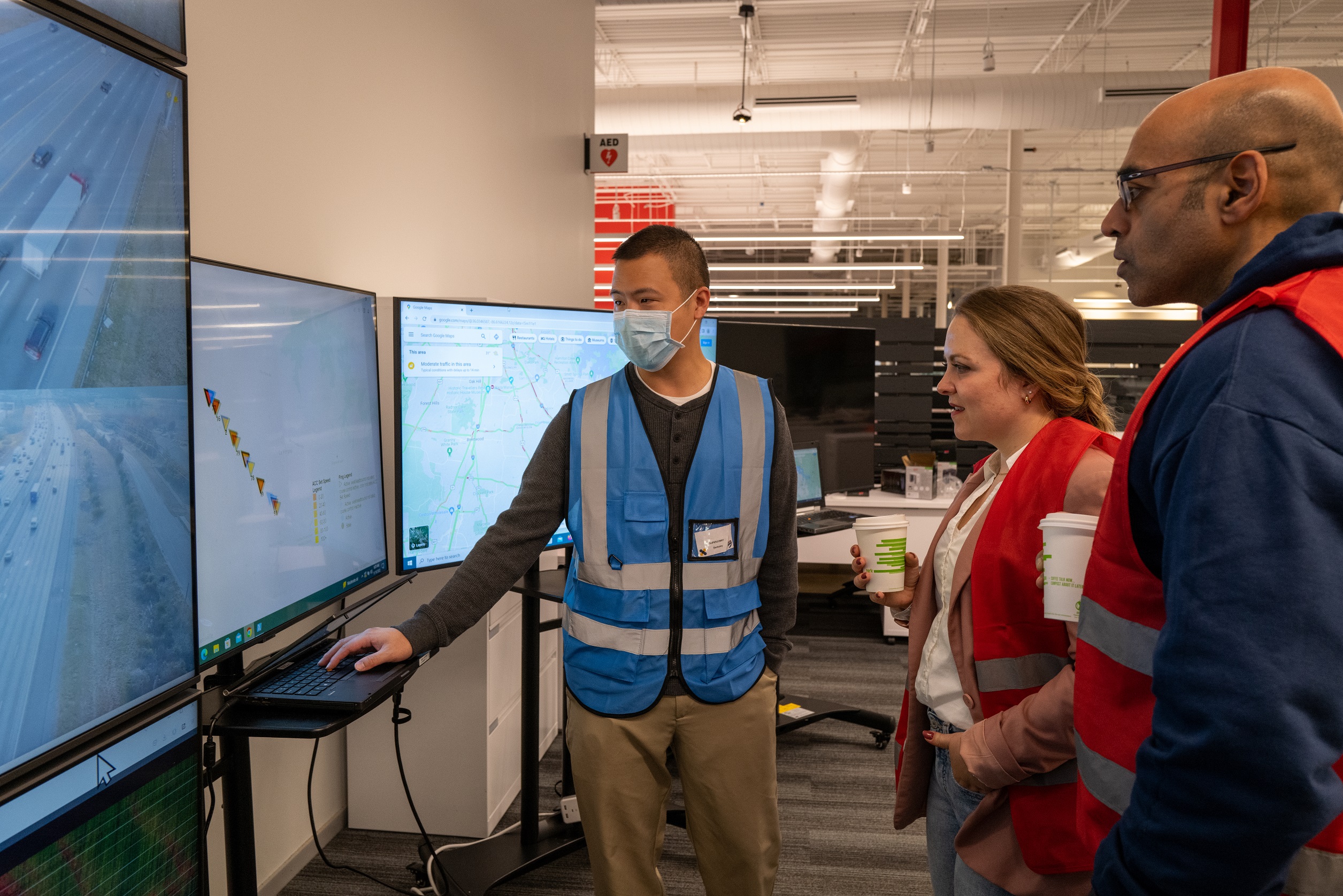}}{The \textit{Field Headquarters} (FHQ) could anticipate drivers returning by tracking the vehicles. Dr. Jonathan Lee (left), a designer of the monitoring system of the AVs from the FHQ, is showing the system to the U.S. Department of Energy (DOE) representatives, Heather Croteau (center) and Prasad Gupta (right) from the \textit{Vehicle Technologies Office} (VTO). \label{fig:JonnyVisScreen}}

    The hardware team devised a plan to validate that all of these components during the full experiment lifecycle.  Each Raspberry Pi had a script installed named piStatus which compiled a set of status information and pushed the full status to a server.  piStatus was built to monitor states provided by libpanda and a middleware support software suite named can\_to\_ros to provide live vehicle states \cite{elmadani2021can,nice2023middleware}. A basic web client then sorted this information for the ability to rapidly identify issues.  Color coding was also used to identify if a system had a problem (red), was busy powering up (yellow), recently becoming offline (orange), was all clear (green), or offline (grey).  In summary, piStatus provided the following live information:
    \begin{itemize}
        \item Network Status - Connection status and Datarate
        \item System Power
        \item Software Version
        \item Hardware Status
        \item Hardware Live Checking
        \item Raspberry Pi MAC to Vehicle Number Mapping
    \end{itemize}

\sdbarfig{\label{fig:pistatus-checking}\includegraphics[width=19.0pc]{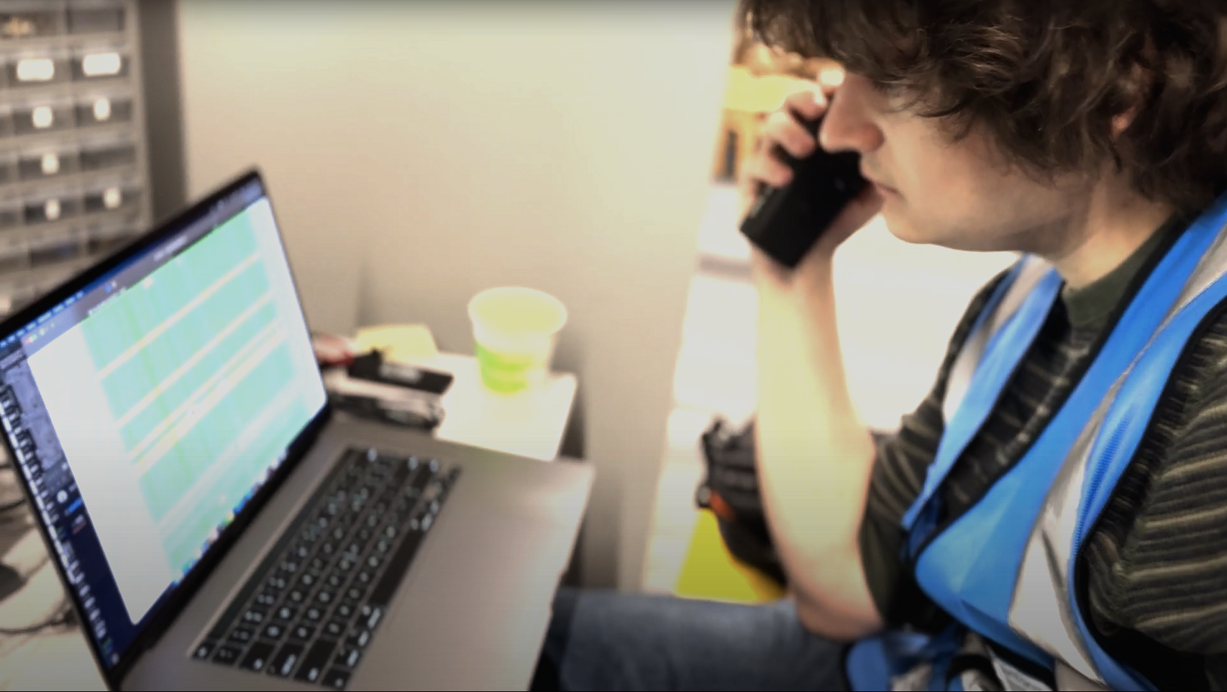}}{During each experiment, headquarters personnel (Matt Bunting shown here) closely monitored vehicles from vehicle start through vehicle shutdown to note which vehicles were ready for deployment or required later maintenance.}

    % \sdbarfig{\label{fig:pistatus-snippet}\includegraphics[width=19.0pc]{figures/pistatus-cropped.png}}{ Derek Gloudemans (left) and Matt Bunting (center) monitor the piStatus of the AVs.}

    The information provided by piStatus was invaluable for the success of experiment deployment.  Figure~\ref{fig:pistatus-checking} shows an operator monitoring piStatus and clearing vehicles for deployment by ensuring all states were green.  piStatus would be continuously monitored throughout the experiment and vehicle issues would be noted for performing maintenance before the next experiment.  In the total of four experiments performed, a vehicle operation yield of 100\% was achieved.  %The success of piStatus can be seen in figure~\ref{fig:all-100}.

    % removed fig for space
    % \sdbarfig{\label{fig:all-100}\includegraphics[width=19.0pc]{figures/all-100.png}}{The final vehicle leaves for the experiment in the final experiment day, showcasing a 100\% yield of all hardware installs, only possible by using piStatus.  }
    
\end{sidebar}

\begin{sidebar}{Daily workflow and autonomous vehicle release schedule}
\noindent
by Sean T. McQuade
% {Morning preparations and scheduling}

% \setcounter{sfigure}{8}
\renewcommand{\thesfigure}{S\arabic{sfigure}}

\noindent
\sdbarfig{\includegraphics[width=19.0pc]{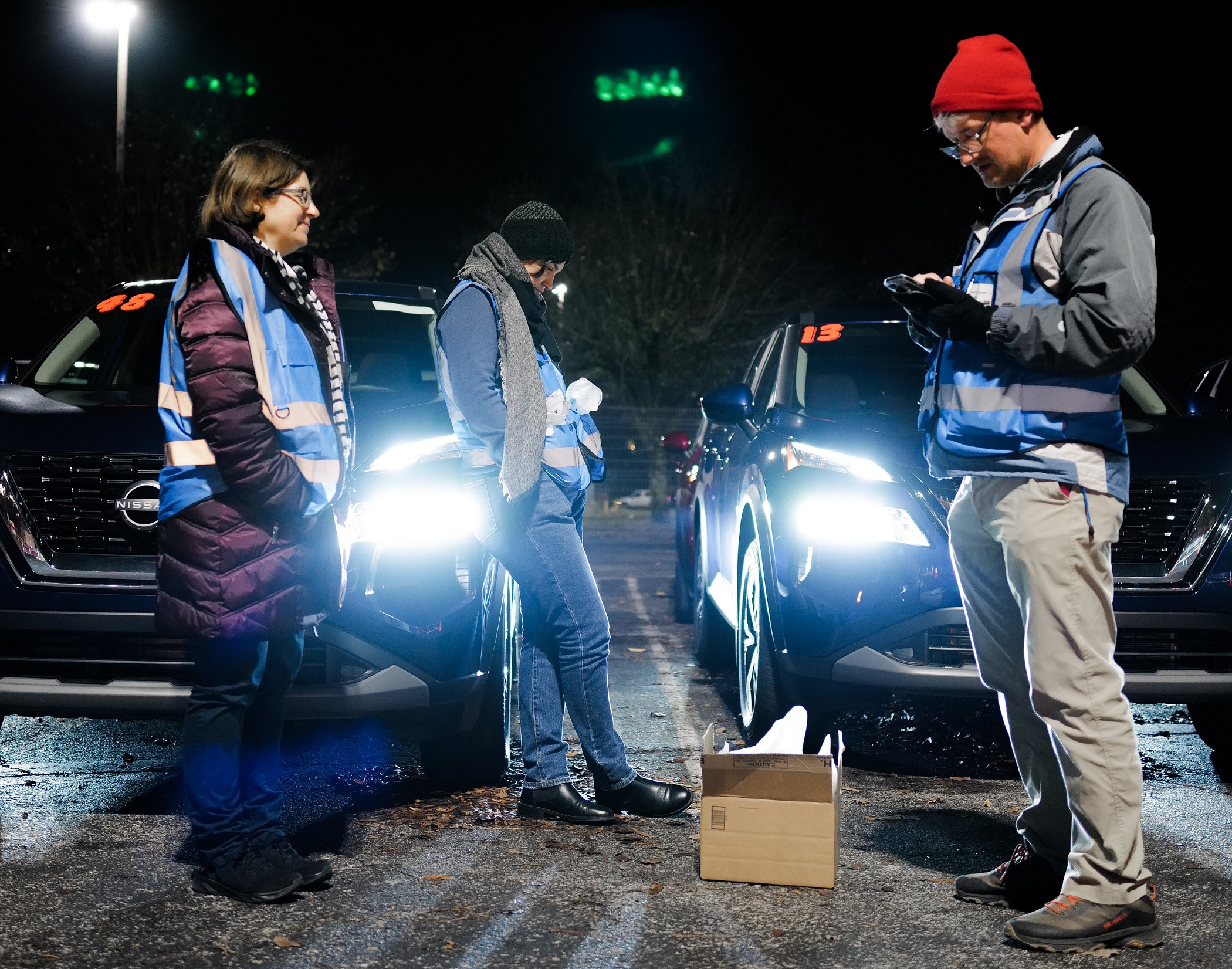}}{From left: Sharon Hornstein, Maria Laura Delle Monache, Jonathan Sprinkle guiding morning preparations.\label{fig:DarkLot}}

% The sign in process to keep drivers organized and ready to enter vehicles as needed. Key management system we used. I will consult Riley about more details.

\sdbarinitial{T}\emph{o} execute the 100 AVs test smoothly in the appropriate traffic conditions, all CIRCLES personnel arrived at the experiment headquarters on Hickory Hollow PKWY by 5:00am.
Drivers were instructed to arrive onsite by 5:30am. Upon arrival, drivers were greeted at the door, passed through a Driver’s License check, then received their name badge with their name, driver ID, and designated route (orange or yellow). Next, they received a vest correlated to their route color with a vest ID sewn into the fabric. The driver ID and vest ID were recorded in the attendance database. Drivers then headed to a daily safety briefing before they were released to the waiting area. Their names would populate a screen showing a queue so that they would prepare to go to the lobby area so that they could quickly exit the building and come to the \textit{Field Headquarters} (FHQ) parking lot. At the lobby, they would wait to be released by an attendant to the FHQ parking lot area. The first drivers headed out to the AVs at 6:00am so that most vehicles could be on the routes by 7:30am. 

Upon arrival in the parking lot, a key corresponding to the driver’s assigned route would be checked out and given to the driver. The driver walked through the pedestrian area to their vehicle. With assistance from the parking lot crew, the drivers prepared their route to be displayed on the screen in the center console of the vehicle. A crew member would talk through the steps to engage the cruise control and remind the driver of their driving route. The crew member would also point out emergency contact phone numbers located in the vehicle.

The controllers had been installed on a Raspberry Pi that was connected to the CAN bus. To ensure the installed controller would run correctly, parking lot staff had to verify the Raspberry Pi was turned on after the driver started the vehicle. If it remained off, we would power cycle the pi to confirm it was on.

When the driver was prepared to go, they would receive a final briefing from Professor Dan Work or Professor Jonathan Sprinkle. This was to remind drivers of their instructions and safety protocol. They ended by asking the driver if they felt comfortable and if they had any questions.
The drivers were then told to follow the ``flow marshall's'' instructions. Flow marshalls were a parking lot crew of eight members responsible for guiding the vehicle out of the parking lot safely with lighted batons.

% \noindent
% \sdbarfig{\includegraphics[width=19.0pc]{figures/GantChart.png}}{An idealized schedule of the first 100 drivers and their breaks. It shows each driver assuming they maintain the order from which they departed the headquarters. This was used to estimate the number of extra drivers needed to keep all vehicles on the road during breaks. AVs were assigned lanes. The orange/yellow segments indicate orange/yellow route drivers driving Westbound, the gray segments indicate them driving eastbound back to start another loop, and the blue segments indicate drivers returning to the lot for a break.\label{fig:Ganttchart}}
\end{sidebar}

\section{Dynamic agent-based simulation framework}
This section presents the creation and calibration of a high-fidelity microscopic simulation scenario using available data on the I-24 road network (Figure \ref{fig:I-24}). The goal is to generate a realistic scenario and introduce AVs before the experiment in order to monitor the traffic condition and verify the experiment scenario by simulation. 

\subsection{Dynamic traffic simulation}
In the realms of both academic inquiry and practical application, dynamic traffic simulators and \textit{Dynamic Traffic Assignment} (DTA) models are noteworthy tools for replicating real-world traffic conditions and forecasting the impacts of novel policies or infrastructure changes. These models are underpinned by a multitude of parameters and decision variables, which are essential in faithfully capturing the dynamics of actual traffic \cite{keimer2020routing}. Proper calibration of these variables, informed by real-world data, is vital to align the simulation's outcomes closely with observed traffic patterns.

In the initial phase of the study, a comprehensive literature review was conducted, focusing on both macroscopic and microscopic traffic models \cite{seo2017traffic}. These models offer a foundation for developing a comprehensive method to gauge the effects of integrating AVs into existing traffic networks. Summaries of these models are presented in "Microscopic Traffic models" and "Macroscopic Traffic models" for ease of reference. However, to secure both the safety of the AVs and a detailed representation of their interactions with human-driven vehicles, we employ agent-based traffic simulation. This method accommodates a granular level of detail, encompassing road infrastructure, traffic signals, and even lane-changing behaviors. Note that the AVs do not change the lane in the highway in the experiment zone described in "AVs Deployment in MVT Test"

The use of an agent-based approach \cite{bazzan2014review} allows us not only to understand the present state of traffic but also to assess how the incorporation of new vehicles might exacerbate or alleviate congestion. This article delineates the methodology we employed to calibrate this sophisticated, simulation-based DTA framework, thereby enhancing the reliability and applicability of our findings in the domains of transportation planning and policy evaluation.

%this simulation was not used to test controllers or train RL algorithms for controllers. 

% \begin{figure}[h!]
%     \centering
%     \includegraphics[scale = 0.26]{figures/Sumo_I24.png}
%     \caption{I-24 microsimulation network}
%     \label{sumo:network}
% \end{figure}

The required inputs for a microsimulation are a network (presented in Figure \ref{sumo:network}), dataset \textit{origin-destination} (OD) demand, and the demand distribution over time and paths \cite{hollander2008principles}. The traffic data are used to calibrate and validate the simulation. TDOT provides us with a flow data set (\#vehicles/hour, see Figure \ref{fig:TDOT_I24_volume}), the \textit{Level of Service} (LOS) data, and the cycle time of signals. LOS data provide us with data regarding the level of flow for the targeted speed considering the characteristics of the road, for example, road type, standard road width, and number of lanes \cite{prassas2020concepts}. In addition, the speed data is provided by INRIX. %TDOT flow data corresponds to the network of roads between intersections (i.e., network link) collected every hour, while the INRIX data gives the speed for every road segment collected every minute. One network link can consist of more than one road segment, and there is no overlap between road segments and links. 
Both data sets represent the morning peak hour (6:30am - 8:30am). 
The heterogeneity in data aggregation and temporal resolution presents a significant challenge in this study. Specifically, TDOT data, sourced from loop detectors, furnishes average hourly flow metrics for an entire year and corresponds to each edge of the network graph. On the other hand, INRIX data, collected from Probe vehicles, provides average speed readings on a minute-by-minute basis for individual days and is associated with road segments, each typically half a mile in length. Consequently, each network edge can be seen as a composite of multiple road segments. For further information regarding the Probe vehicle, collecting traffic data, and traffic monitoring, please refer to \cite{10.1145/1378600.1378604, claudel2010linear, tinka2010quadratic, seo2017traffic}.

% A short literature review can be added.

To reconcile these divergent spatiotemporal characteristics and decision variables across the two data sets, we employ bi-level programming to formulate the calibration problem, wherein each level targets a specific variable for calibration. In this article, we initially discuss the construction of the network and the development of the simulation framework. Subsequently, we review the state-of-the-art on DTA calibration and introduce the proposed methodology for calibrating these disparate data sets to create a simulation-based framework that allows us to rigorously test various scenarios for the real-world experiment.

\begin{sidebar}{Microscopic traffic models}
\noindent
by Mostafa Ameli and Benedetto Piccoli

    \sdbarinitial{M}\emph{icroscopic traffic models} aim to represent the movement of individual vehicles. These models use multiple coupled \textit{Ordinary Differential Equations} (ODEs) (one per vehicle) to determine each vehicle's position, speed, and acceleration. These ODEs compare the position or speed of a given vehicle with the one just in front, its Leader, to get the acceleration of the first vehicle
    \cite{Gazis61}. For instance,  the difference in position is often referred to as a space gap. 
    Multi-class models include different types of vehicles (for example cars and trucks), while multi-lane requires modeling of lane-changing maneuvers (see \cite{Xiaoqian2023}).

    If we have $N$ vehicles on a single lane, which we order from first to last (vehicles cannot overtake each other in this model so they always have the same rank). The idea is to determine the speed of every vehicle $i$ based on the leading vehicle $i-1$'s speed characteristics. Therefore, each vehicle tries to match its speed with the leading vehicle's speed:
    \begin{equation}
        \begin{split}
            &\dot{x_i} = v_i\\
            &\dot{v_i} = \beta_i \frac{v_{i-1}-v_i}{(x_{i-1}-x_i)^2} \;,
        \end{split}
    \end{equation}
    where $x_i$ is the position and $\dot{x_i}$ the speed of vehicle $i$. The leader vehicle of the platoon is indexed $1$ and usually follows the dynamics:
    \begin{equation}
        \begin{split}
        &\dot{x}_1 = V_{\mbox{max}}\\
        &\dot{v}_1 = 0.
        \end{split}
    \end{equation}
In the Bando model \cite{Bando95}, each vehicle tries to match an optimal speed determined by its distance to the vehicle immediately in front:
\begin{equation}
    \begin{split}
        &\dot{x}_i = v_i\\
        &\dot{v}_i = \alpha_i (v_i-V(x_i-x_{i-1}))\;,
    \end{split}
\end{equation}
where $V(\Delta x)$ is the optimal velocity (which is positive) associated with the space gap $\Delta x$,
%This should be close to 0 if $x$ is big; 
%We expect it to be an increasing function.
%For instance, we can take: 
for instance $V(\Delta x)=\tanh(\Delta x)$.
%The position and the speed of the leader $1$ are still defined by:
%\begin{equation}
%    \begin{split}
%        &\dot{x}_1 = V_{\mbox{max}}\\
%        &\dot{v}_1 = 0
%    \end{split}
%\end{equation}
We can combine these two models to get the Bando-Follow-The-Leader \cite{delle2019feedback, Gong2023}:
\begin{equation}
    \begin{split}
    &\dot{x}_i = v_i\\
    &\dot{v}_i = \alpha_i (v_i-V(x_i-x_{i-1}))+\beta_i \frac{v_{i-1}-v_i}{(x_{i-1}-x_i)^2}\;.
    \end{split}
\end{equation}
Here the parameters $\alpha_i$ and $\beta_i$ are positive numbers that may depend on the type of vehicle $i$ (truck, car, ...).
Finally, the Intelligent Driver Model \cite{albeaik2022limitations,IDM00} is given by:
\begin{equation}
    \begin{split}
    \dot{x_i} &= v_i\\
    \dot{v_i} &= a\times (1-(\frac{v_i}{v_m})^{\delta}-(\frac{s^{*}(v_i, v_i-v_{i-1})}{x_{i-1-x_i-l_i}}^2)\;,
    \end{split}
\end{equation}
where $l_i$ is the length of vehicle $i$,
$s^{*}(v, \Delta v) = s_m+v\times T+\frac{v\delta v}{\sqrt{2ab}}$ while $a, b, T, s_m, v_m, \delta$ are parameters of the model. Each driver modulates their speed according to the space gap, and the difference in speeds with the vehicle in front.

Furthermore, to take into account the presence of control vehicles such as in the \textit{MegaVanderTest} (MVT) experiment, one could divide the vehicles into two categories (1) vehicles subject to the previous system, and (2) control vehicles whose speed would be prescribed by an algorithm and depend on the state of traffic (position, time, and the state of other vehicles) \cite{Hayat2023theory}.
\end{sidebar}

\subsection{Network creation}
Based on the purpose of this study, within the traffic simulation options, only dynamic microsimulations can
help understand the impact of adding AVs on road traffic at the scale of the I-24 network. Within
the available traffic microsimulator, we use SUMO as an open-source simulator motivated by previous projects \cite{langer2021calibration}. In SUMO simulation, each driver (agent) has multiple decisions to make, for example, route, lane, and departure time choice. 

The road network can be represented as a graph consisting of constituent road sections (links or edges) connected through signalized or unsignalized intersections (nodes). To create the I-24 network in SUMO, the mapping data was imported from the OpenStreetMap \cite{haklay2008openstreetmap}. The road network consists of 154 nodes and 452 links. The network's topology and physical characteristics are verified with multiple mapping data, including Google Maps, Bing Maps, and Apple Maps. The verification carried out on the I-24 SUMO network presented in Figure \ref{sumo:network} results in various modifications, such as modifying edge lengths and the number of lanes. Figure \ref{sumo:intersections} presents four intersections of the I-24 network after the verification using mapping data. In controlled junctions, 16 traffic signals were also modeled based on TDOT data that determined the cycle time of signals.

\setcounter{figure}{5}
\begin{figure*}[!]
    \centering
    \includegraphics[width=\textwidth]{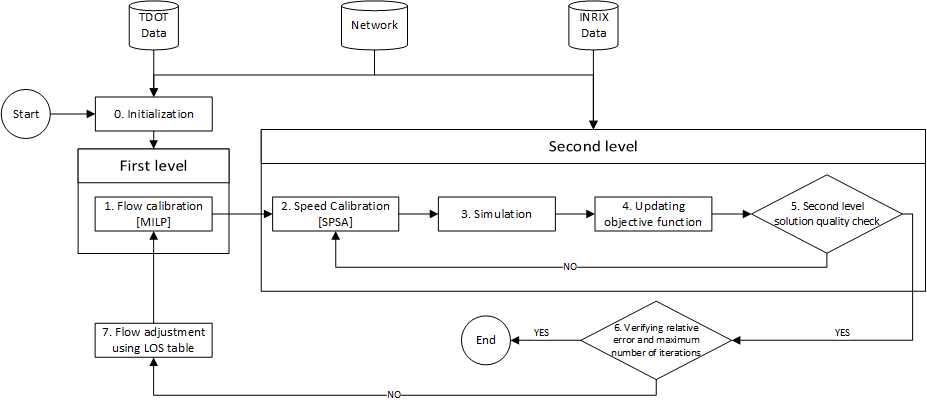}
    \caption{Bi-level DTA calibration framework: Step 1 focuses on calibrating flow by minimizing the \textit{Mean Squared Error} (MSE) between actual and reference link flows, resulting in the \textit{Origin-Destination} (OD) matrix and the count of agents traversing specific OD pairs. Transitioning to Step 2, the emphasis is on determining optimal agent departure times for speed calibration, employing a modified algorithm with a preliminary assumption of a uniform distribution for these times. Step 3 initiates a simulation, with a 20-minute warm-up to integrate agents smoothly, during which the average speeds of road segments are measured and compared to Probe vehicle data. Step 4 then adjusts the objective function for speed based on these simulation results. Step 5 evaluates algorithmic convergence using set criteria; if not met, the process revisits Step 2. Step 6 assesses overall solution quality, and if any link surpasses a 10\% absolute error, it proceeds to Step 7. This final step addresses temporal data mismatches by adjusting values with reference to the network's \textit{Level of Service} (LOS). If resultant errors are deemed minor and speed calibration meets the threshold, the model is considered converged; otherwise, the cycle returns to Step 1 for further refinement.}
   \label{fig:flow chart}
\end{figure*}

\subsection{Supply and HDV driving model calibration}

\setcounter{figure}{5}
% \begin{figure}[h!]
%     \centering
%     \includegraphics[trim={1cm 0cm 2.6cm 0}, clip,scale=0.23]{figures/ParkingLotFlow.PNG}\\
%     %\includegraphics[trim={0 6cm 0 4cm 0}, clip, scale=0.42]{figures/Lot Photo.jpg}

%     \vspace{0.1cm}
    
%     \includegraphics[trim={3cm 6cm 1cm 4cm 0}, clip, scale=.565]{figures/Lot Photo.jpg}
    
%     \caption{Flow of experimental vehicles through our lot. Top: The planned positions of vehicles and crew. This was designed to allow the drivers of two different routes (orange and yellow) route  to safely check in keys and substitute for each other. Bottom: A photo of the lot, orange route vehicles on the left, yellow on the right.}
%     \label{fig:Lot layout}
% \end{figure}

Determining each vehicle's departure time and the route is equivalent to assigning the OD demand to dynamic routes, which results from DTA models \cite{ameli2020cross}. Once the departure time and route of each vehicle are known, the simulator computes the dynamic road section traffic loads in the network \cite{ameli2020simulation}. One of the crucial steps to achieving realistic results from simulation tools is calibration. It aims to determine the DTA model's input such that the output represents traffic scenarios with a reliable level of accuracy \cite{antoniou2004line, ameli2021computational}. The inputs can be divided into two groups: demand and supply. The supply parameters define the environment of the simulation and the field constraints, for example, traffic network topology and capacity, traffic signals, and speed limitation. In contrast, the demand inputs represent the travelers and their behavior in the system, for example, time-dependent origin-destination matrix, routing, and lane changing. Based on the available data on link flow and road segment speed, we formulate a new bi-level optimization framework to iterate between two levels to calibrate the simulation scenario with respect to both data sets while considering the correlation between speed and flow results from the agents' route and departure time. Regarding the driving model, we deploy the \textit{Intelligent Driver Model} (IDM) in this study \cite{albeaik2022limitations}. The calibration of IDM parameters is carried out based on the characteristics of the test case, for example, driving laws and culture in the location of the test case. We address this issue by conducting multiple sensitivity analyses on the parameters of the simulator driving model, including (i) speed limit of each edge, (ii) speed factor that lets vehicles exceed the speed limit, (iii) the eagerness to perform strategic lane changing, (iv) the eagerness to perform lane changing to gain speed, (v) the willingness to perform cooperative lane changing, (vi) the eagerness for following the obligation to keep right, (vii) the probability for violating rules against overtaking on the right, (viii) ``lookahead'' time in seconds for anticipating slow down, and (ix) Factor for cooperative speed adjustments. In 2021, a preliminary experiment with 10 vehicles was conducted by the CIRCLES Consortium to evaluate the testbed of the MVT, and multiple trajectories were collected, which led us to conduct a sensitivity analysis on the mentioned parameters and calibrate the HDVs in simulation.

\section{Methodological framework to generate the background traffic}

The existing literature on traffic model calibration often focuses on a singular data type such as flow, speed, or travel time, with many studies employing a weighted sum objective function to handle multiple variables (see for example \cite{seo2017traffic, frederix2011hierarchical,tympakianaki2015c,djukic2017modified,zhang2017efficient,zhu2018calibrating,zhu2019dynamic,gonzalez2021alicante,paz2015calibration,cobos2016calibration,amirjamshidi2019multi, ameli2019heuristic, he2021validated,paz2020calibration,antoniou2015w, claudel2008guaranteed}). Some have utilized the Pareto front method to explore potential solutions, although identifying the optimal solution remains a challenge \cite{cobos2016multi, osorio2015metamodel, cobos2020multi}. Other research has attempted simultaneous calibration for flow and speed, but these are sequential and lack iterative feedback, making optimizations on these systems susceptible to getting stuck in local optima \cite{bachechi2020using,hu2017sequential,osorio2019dynamic,zhu2021joint}. These studies commonly use data from a single source and pre-process it to obtain identical spatiotemporal characteristics, thereby not fully capturing the complexity of real-world traffic networks.

This study aims to advance the existing methodology for traffic calibration by proposing a bi-level mathematical model for agent-based DTA simulation to calibrate both speed and flow based on collected data with different spatiotemporal characteristics. This holistic approach addresses the inherent disparities between data sets and allows for a more nuanced calibration. The methodology employs a feedback loop that facilitates iterative refinement of the calibration process, thereby enhancing the model's accuracy and comprehensiveness. 

Figure \ref{fig:flow chart} presents the bi-level calibration framework proposed to calibrate the simulation of background traffic. The process starts by importing the link flow data and building the network graph. The reference link flow distribution in iteration $m$ of the optimization is denoted by $\hat{X}^m$, representing the number of vehicles passing through each link. For $m = 0$ it is the set of time-dependent link flows collected by the TDOT loop detectors. $\hat{x}^k_{i}$ represents the reference flow of link $i$ ($i \in E$, set of all links) at time interval $k$, $\hat{x}^k_{i} \in \hat{X}^m$. $K$ is the set of time intervals for the flow data, $k \in K$. Note that for the speed data, we consider the same time horizon to address the morning peak. However, the set of time intervals for speed data is different and denoted by $R$ and indexed by $r$, $(r \in R)$. The reason is that the duration of time intervals in $R$ and $K$ is given by the data set and is not equal for both sets. In this study, the flow data is collected every hour, and speed data is collected every minute. Thus, the duration of $r$ is less than $k$. In the first (upper) level, the flow calibration problem is formulated as a \textit{Mixed Integer Quadratic Programming} (MIQP) model and solved to determine the path flow distribution, $\Pi$. $\Pi$ determines the number of agents who start their trip on each path at each time interval. The flow of link $i$ at time $k$ resulting from $\Pi$ is denoted by $x^k_i$. 

The objective function of the upper level (Step 1 in Figure \ref{fig:flow chart}) is to minimize the \textit{Mean Squared Error} (MSE) between $x^k_i$ and $\hat{x}^k_{i}$. The output of this step is the OD matrix and the number of agents traveling on paths between each OD pair, $\pi_{OD}, for all \pi_{OD} \in \Pi$. We fixed the path flow distribution for the second (lower) level, wherein we determined the departure time distribution of agents to calibrate the speed variable. To find the optimal departure time, we propose a modified version of the \textit{simultaneous perturbation stochastic approximation} (SPSA) algorithm inspired from \cite{lu2015enhanced}. For the initial solution in Step 2, we consider the uniform distribution for the departure times. In every iteration of the SPSA algorithm, we run a simulation (Step 3) with the demand profile from the previous step. Note that 20-minute simulation warm-up is considered at the beginning of the simulation to insert the agents with their optimal departure times. In Step 3, We also measure $s_l^r$, the average speed of road segment $l$ ($l \in L$, set of all road segments in INRIX data) at time intervals $r$ and compare it to the corresponding value $\hat{s}_l^r$ from data set $\hat{S}$ collected by the Probe vehicles. %\footnote{Probe vehicle: a vehicle with an onboard data collection devices that observes traffic conditions and measure related indicators while the vehicle is in the traffic flow.}. 
Then, in step 4, the objective function (MSE of speed variable) of the lower level is updated based on the simulation results. 

The next step is checking the convergence conditions of the SPSA algorithm based on the maximum number of iterations for the lower level and comparing the solution quality (MSE of the speed values) with a threshold. If the convergence is not achieved, we go to Step 2. Otherwise, we perform the second level solution quality check in Step 6 by considering the flow and speed error distribution and the maximum number of iterations for the bi-level framework. If there is any link with more than 10\% absolute error, we go to the next step; otherwise, we finish the process. In Step 7, we aim to address the temporal correlations between two data sets. We modify the values in $\hat{X}$ with respect to the level of service (LOS) of the targeted network. LOS gives us a level of flow for the targeted speed considering the characteristics of the road \cite{prassas2020concepts}. For example, if the speed measured by simulation is greater than the data, we increase the value of the target flow with a step size function $g(\cdot)$. As a result, the density will increase at time $k$, and then we can expect the speed to reduce at $r$. This modification results in additional errors for the upper level as the targeted flow value is modified ($\hat{X}$). Therefore, in Step 
7, we check the relative error of flow and speed in addition to the convergence of the model by the maximum number of upper-level iterations. In other words, if the modification of the flow is minor and the speed error is acceptable, we converge; otherwise, we go to Step 1 to update the path flow distribution.

To clarify the optimization model behind this framework, the bi-level mathematical model used in Figure \ref{fig:flow chart} is as follows:

\begin{align} %added & to first row to fit to one col
\min_{\pi_j}& \; f(x^k_{i},{\hat{x_{i}}^k})=\frac{1}{|E|}{\sum_{\forall k} \sum_{\forall i} (x^{k}_{i}-{\hat{x_{i}}^{k}})^2}\label{eq5} \\\
\notag s.t.\\
& x^{k}_{i}=\sum_{\forall j} \lambda_{ij}\pi^{k}_{j}, \quad \forall i,k,j;  \label{eq6}  \\
% & \hat{x_{i}}^{k}=\begin{cases}Q(LOS,g(.))\;\;\;\;\;\;\;\; \;m>1\\
% TDOT\;\;\;\;\;\;\;\;\;\;\;\;\; m=1\\ 
%  \end{cases}  \label{eq65}\\
& x^{k}_{i}\geq 0, \hat{x}^{k}_{i} \in \hat{X}^{m}, \pi^{k}_{j}\geq 0, \pi^{k}_{j} \in \Pi^{m}; \quad \forall i,k,j.  \label{eq7}
% & k \in K; \quad \forall k \\
% r \in R; \quad \forall r
\end{align}
\\
\begin{align}
\min_{I^{k}_j}\;f(S^{k}_{l},{\hat{S_l}^{k}})=\frac{1}{|L|}\sum_{\forall k}\sum_{\forall r} \sum_{\forall l} {(S^{k}_{l}-{\hat{S_{l}}^{k}})^2}\label{eq8}\\
\notag s.t.\\ 
S^{k}_{l}=h(\Pi^m, [I^{k}_{j}, \forall j]), \quad \forall l,k; \label{eq9}\\
 \sum_{\forall k} \sum_{\forall r} {I^{k}_{j}}=1, \quad \forall j; \label{eq10}\\
 {I^{k}_j} \in [0, 1], \quad \forall j,k. \label{eq11}
\end{align}

\noindent \textit{Step 7:}
\begin{align}
% \notag \text{\underline{\textit{Step 7:}}}\\\
\hat{X}^{m}=\begin{cases} \text{TDOT data}, & \text{First iteration};\\ 
Q(LOS,g(\cdot)), & \text{otherwise}.\\
 \end{cases}  \label{eq12}
\end{align}

In equation \eqref{eq5}, $\hat{x}^{k}_{i}$ serves as an indicator for the benchmark flow on link $i$ during time slice $k$. This calibration problem can be cast into an MIQP, which is amenable to solutions by standard optimization solvers. For agent-based simulation calibration, the path flow distribution needs to be specified. The relationship between individual edge flow and overall path flow is captured by constraint \eqref{eq6}, which introduces a binary variable $\lambda_{ij}$ to indicate if link $i$ is a part of path $j$. Recall that the first stage of this approach yields an OD matrix along with the count of agents traversing each path between every OD pair. The lower level of the problem then focuses on time-dependent inflow rates to determine the departure times for these agents.

The lower-level calibration, formalized through \eqref{eq8}, targets the minimization of discrepancies between observed and simulated speed data to determine the inflow share of each path, $I^{r}_j$ for time interval $r$ respect to constraint \eqref{eq10}. Note that this decision variable is defined over the whole period of interest to optimize the departure time distribution of all agents and address the correlation between time intervals $k \in K$. Equation \eqref{eq9} utilizes the function $h(\cdot)$ to extract simulated speed measures, taking both path flow distribution and inflow rates as inputs. This function can essentially be treated as a black box, allowing for the use of any trip-based simulation. Note that equations \eqref{eq7} and \eqref{eq11} are feasibility constraints for decision variables in the first and second levels, respectively. Moreover, the feedback mechanism described by \eqref{eq12} uses initial average flow data from TDOT for its first iteration, subsequently updating these values ($\hat{X}^{m}$) through the $Q(\cdot)$ function based on the LOS tables in the second level. The first iteration of the second level ($m=1$) uses the exact value in the LOS table. Afterward, for the next iterations, if the same edge is selected, $Q$ revises the flow using a gradient method with adaptive step size \cite{zhou2006gradient}. A technical note is introduced to prevent stagnation in specific edge flow modifications, limiting each edge to be modified for a maximum of $5$ consecutive times. Once this threshold is reached, the one before the last edge with the highest error is selected as the objective edge for subsequent modifications. This provides a dynamic way to adapt the target flow values, enhancing the calibration process. Note that after launching the upper level, the iteration for SPSA in the lower level is reset to $m=0$. The result of the proposed methodology is the path flow distribution to represent the background traffic. In other words, A set of agents is determined, including their departure time and path, which a traffic simulator can use. Therefore, with this solution, we can add the control vehicles with predefined paths and measure their impact on the traffic congestion of the I-24 road network. This enables us to evaluate multiple driving routes for AVs calibrated to the INRIX data. 

\subsection{Validation of the proposed model: Numerical results}

\begin{figure}[h!]
    \centering
    \includegraphics[width=0.48\textwidth]{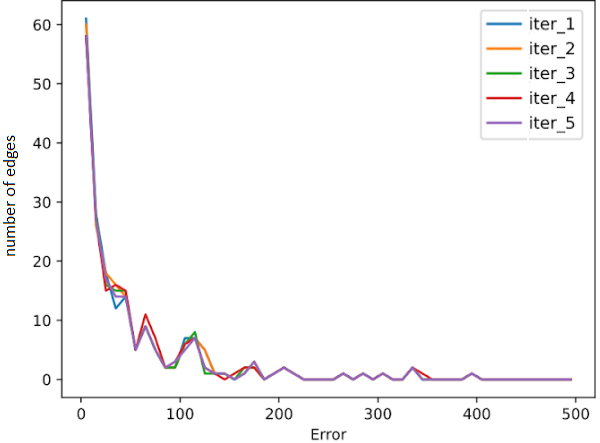}
    \caption{Error histogram of flow calibration: The distribution of links based on their squared error values for link flow, captured through multiple curves. Each curve represents a distinct execution of the initial optimization level, varying by target value and initial solution. The consistency and the small value of error for large number of links across executions underscores the effectiveness of the MIQP solution method employed in our proposed framework.}
    \label{fig:flow-error}
\end{figure}

This section outlines the findings from flow and speed calibration efforts. We used the Scipy optimization solver \cite{2020SciPy-NMeth} to solve the MIQP model at the initial level. Figure \ref{fig:flow-error} displays a histogram of the objective function errors for flow calibration \eqref{eq5} based on five runs of the optimizer. The data indicates consistent performance by the optimization solver, which aligns with expectations for an MIQP-formulated problem.

\begin{figure*}
    \centering
    \subfigure[INRIX data.]
    {
        \includegraphics[width=0.48\textwidth]{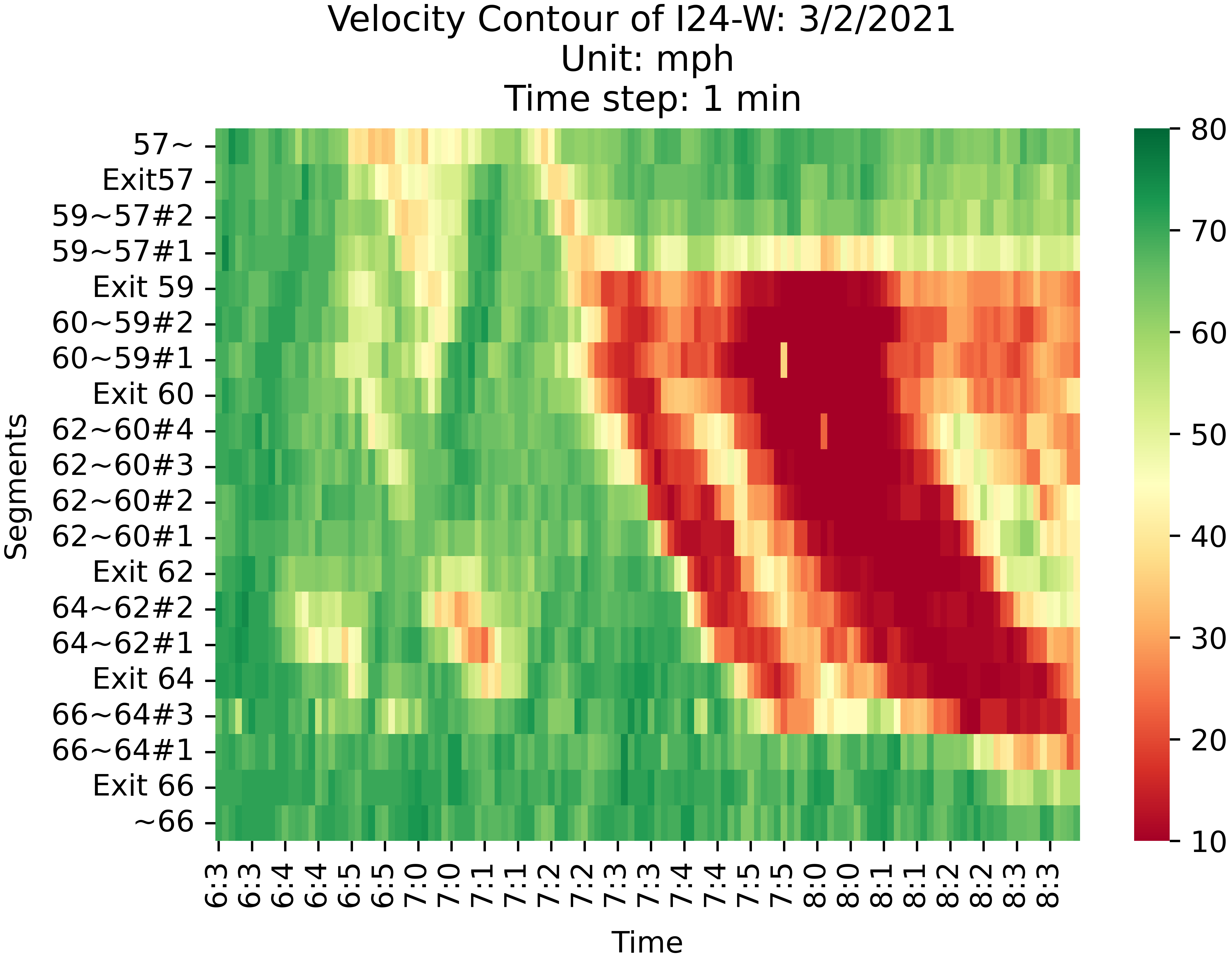}
        \label{fig:Speed HEATMAPa}
    }
    \subfigure[Simulation results with flow calibrator.]
    {
        \includegraphics[width=0.48\textwidth]{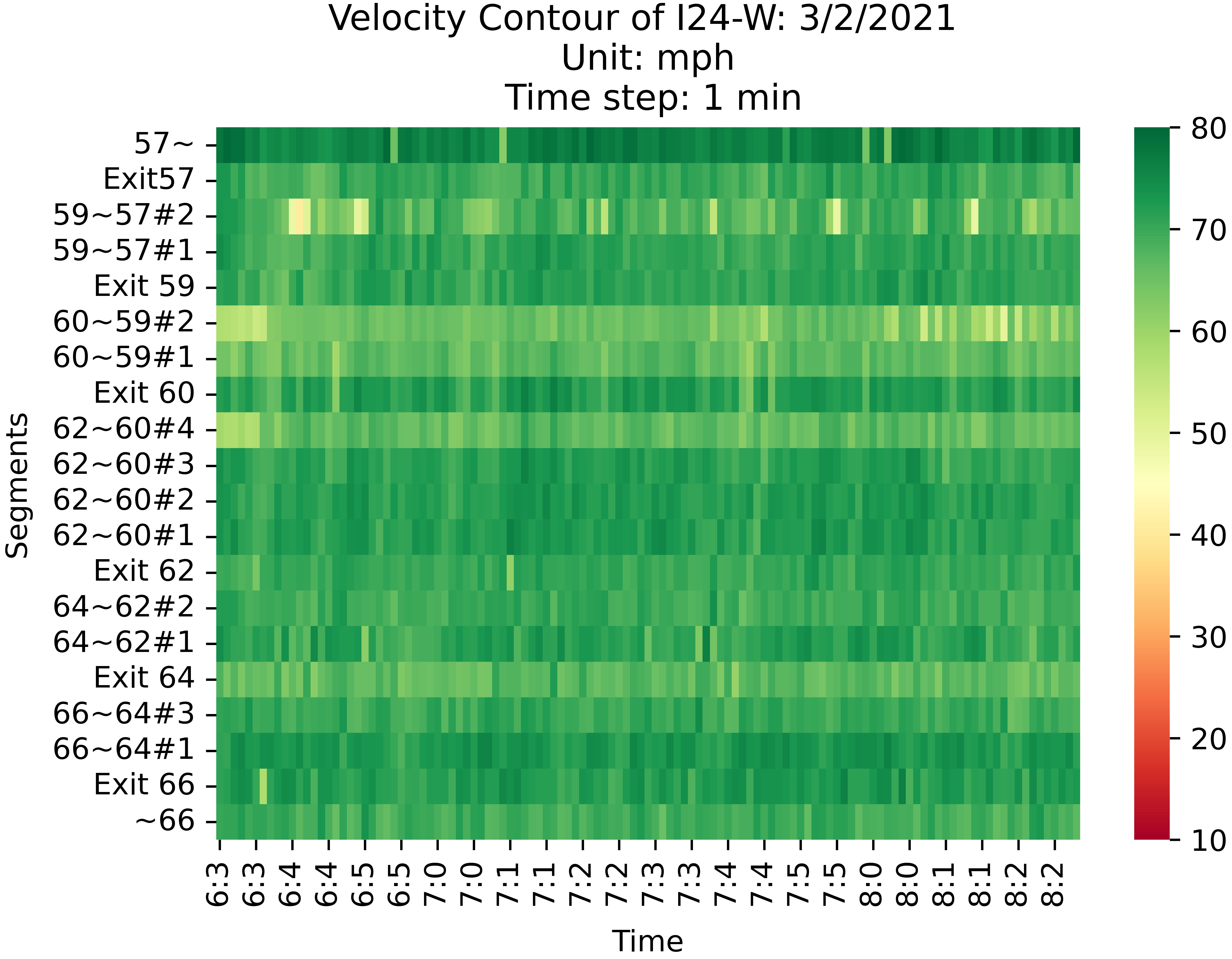}
        \label{fig:Speed HEATMAPb}
    }
    \subfigure[Simulation results with speed calibrator.]
    {
        \includegraphics[width=0.48\textwidth]{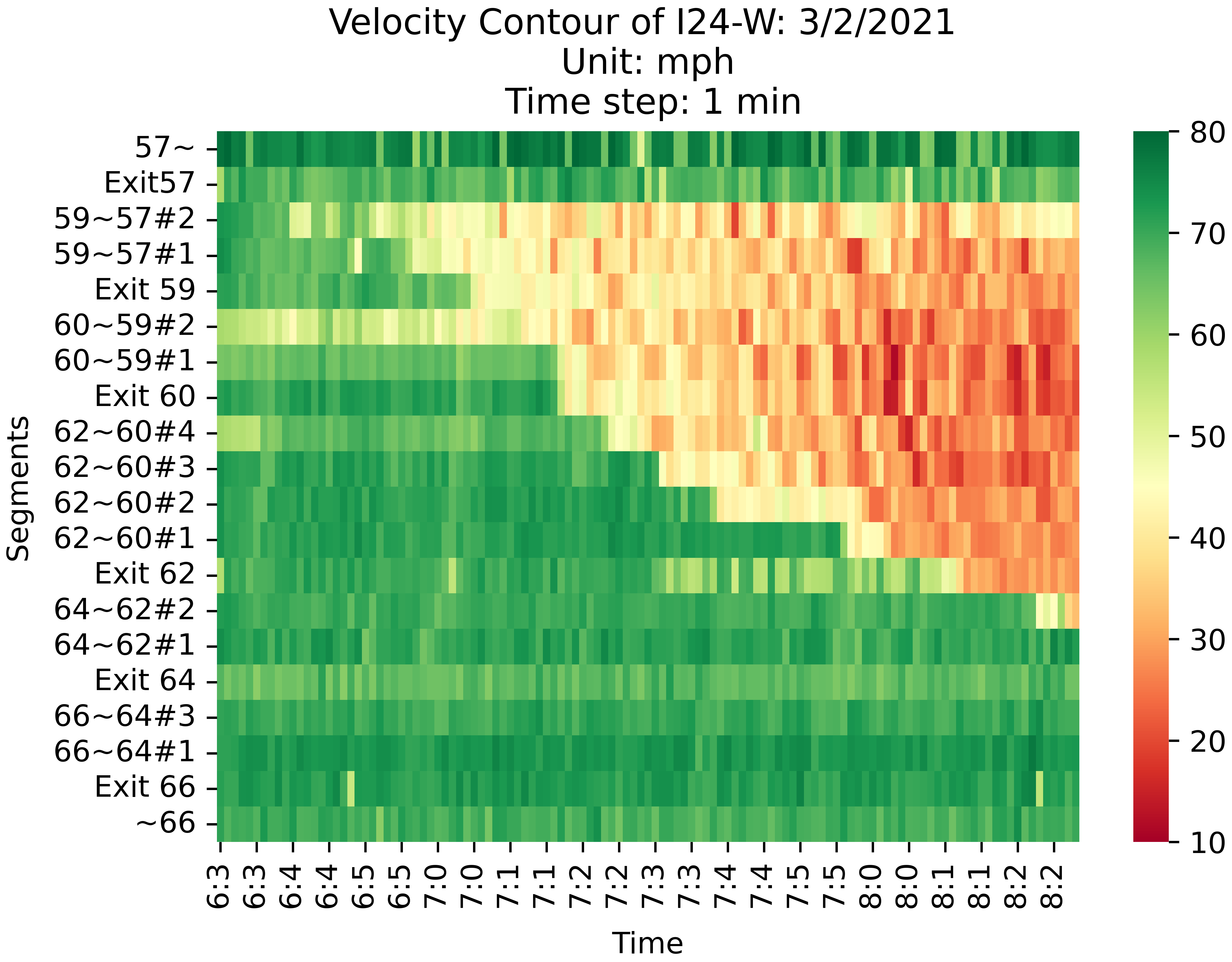}
        \label{fig:Speed HEATMAPc}
    }
    \subfigure[Simulation results with flow and speed bi-level calibrator.]
    {
        \includegraphics[width=0.48\textwidth]{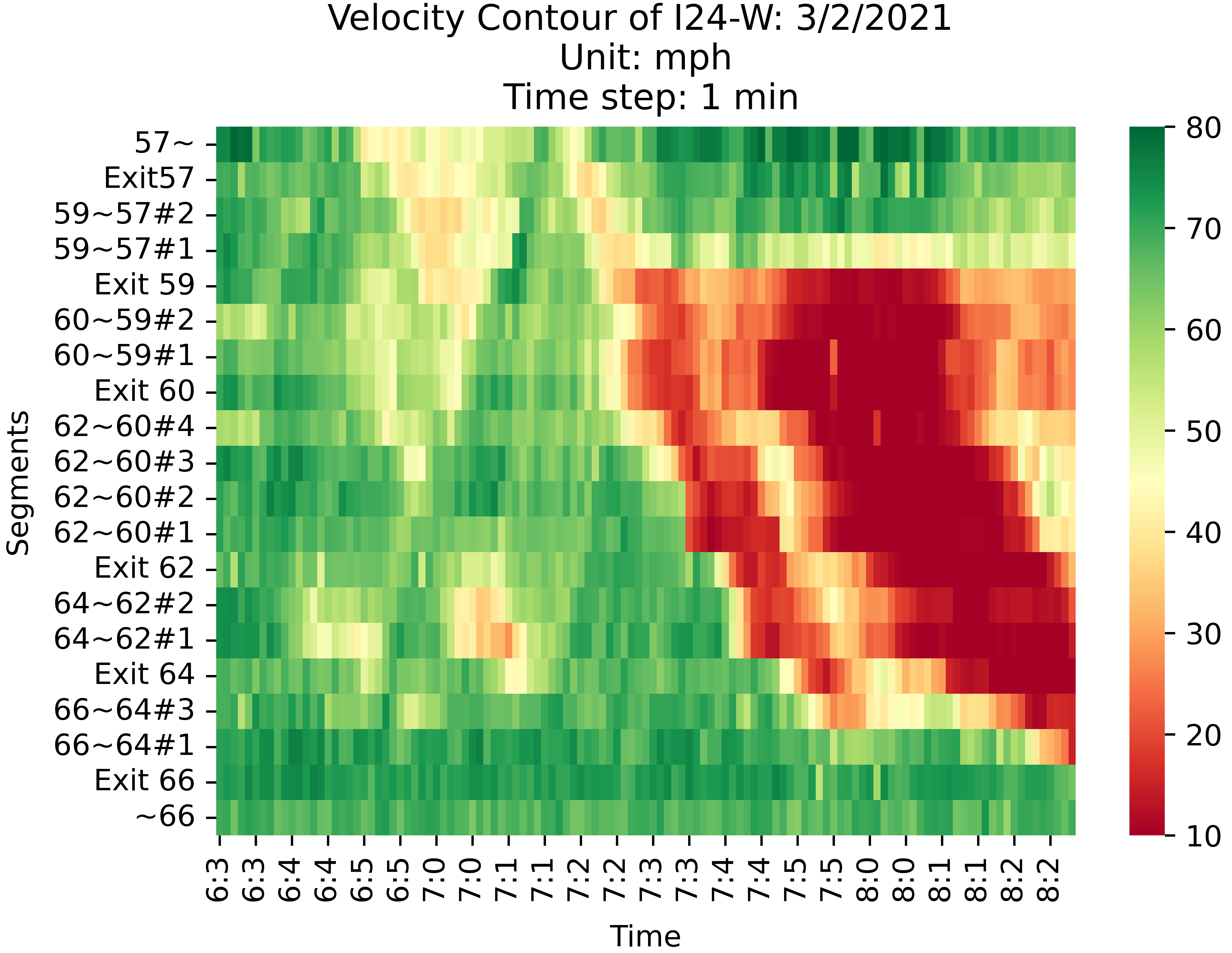}
        \label{fig:Speed HEATMAPd}
    }
    \caption{Spatiotemporal average speed patterns on I-24: Each heatmap displays the speed patterns over various INRIX segments, identified by exit names along I-24, during morning peak-hours (6:30 - 8:30 AM). The y-axis position along the road while the x-axis is marked with time intervals, offering a detailed view of the temporal evolution of traffic speeds. The colored cells indicate average speed values which compare real versus simulated traffic conditions.}
    \label{fig:Speed HEATMAP}
\end{figure*}

We offer an in-depth analysis of the outcomes after running the bi-level code on a single scenario. Figure \ref{fig:Speed HEATMAP} presents average speed patterns across time and space on the road network. The y-axis labels INRIX segments by exit names on the I-24 highway, and the x-axis covers the time range from 6:30 a.m. to 8:30 a.m. The colors indicate speeds (shown on the color bar in MPH). The heat maps show the changes in traffic speeds on various segments of the I-24 highway over time.

Figure \ref{fig:Speed HEATMAPa} depicts INRIX data, featuring a congestion beginning at Exit 59 around 7:00 a.m. Figure \ref{fig:Speed HEATMAPb}, on the other hand, presents simulation outcomes obtained only through flow calibration without using the bi-level method. The simulation indicates an absence of peak-hour congestion, validating the choice to implement the bi-level approach to adjust the speed targets and vehicle throughput.

Figure \ref{fig:Speed HEATMAPc} shows the outcome when using the bi-level approach without feedback. While the scenario attempts to simulate congestion, it does not exactly match the INRIX data patterns. Despite this, the flow is insufficient to induce congestion at Exit 59. Figure \ref{fig:Speed HEATMAPd} presents the effects of incorporating the feedback function with the bi-level method, which resembles the real-world data, including the emergence of stop-and-go traffic patterns.

\begin{figure}[!h]
    \centering
    \subfigure[INRIX data.]
    {
        \includegraphics[width=0.48\textwidth]{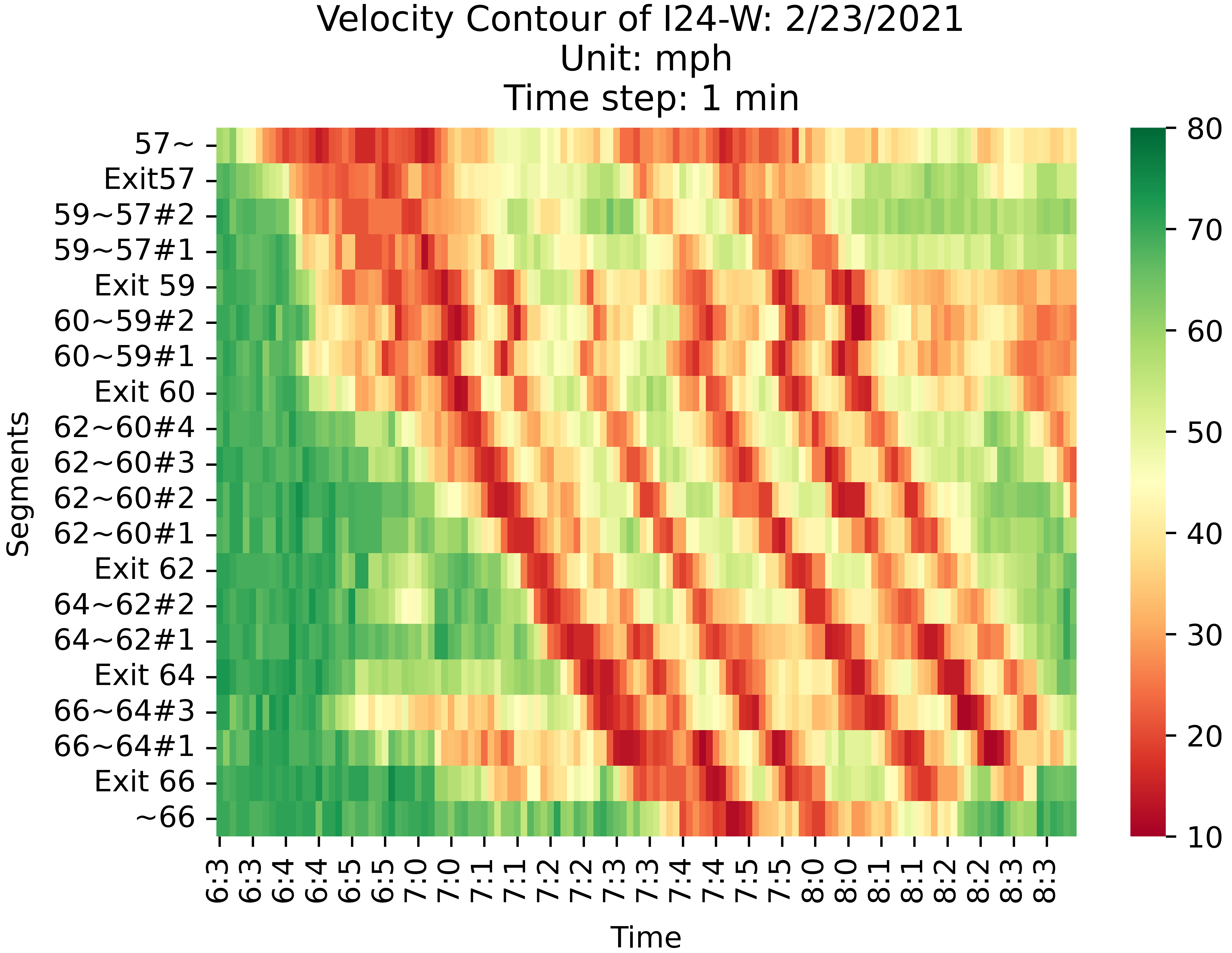}
        \label{Fig:speed patterns comparisona}
    }
    \subfigure[Simulation results.]
    {
        \includegraphics[width=0.48\textwidth]{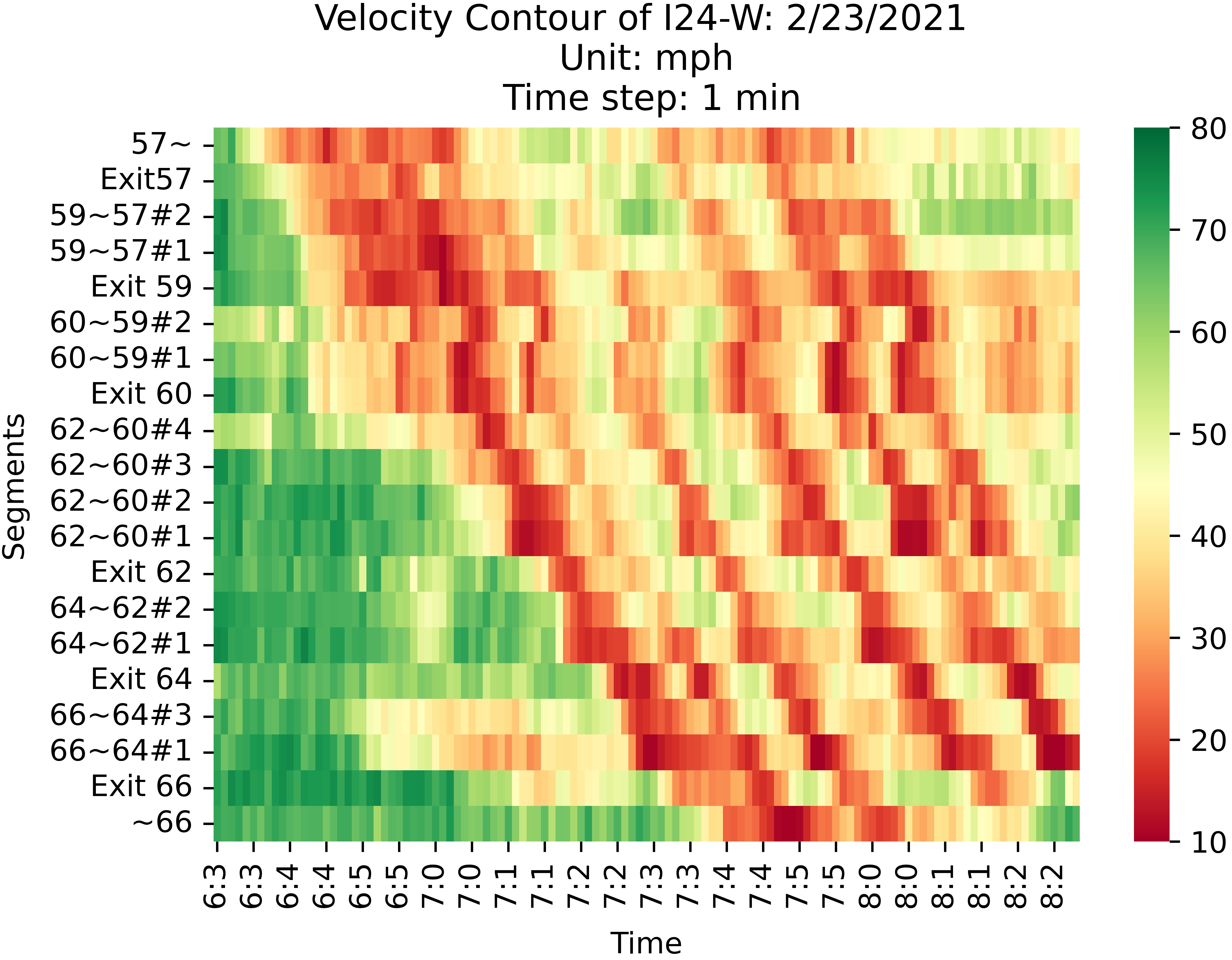}
        \label{Fig:speed patterns comparisonb}
    }
    \caption{I-24 real and simulated Speed patterns: Example of stop-and-go waves.}
    \label{Fig:speed patterns comparison}
\end{figure}

To verify the approach, it was tested on two other INRIX data scenarios. Figure \ref{Fig:speed patterns comparison} illustrates the speed patterns resulting from these tests. The figures show that the simulated outcomes are similar to the real-world stop-and-go patterns, although the simulation shows smoother speed changes compared to the real data. In both mentioned scenarios, we measured the \textit{Root Mean Square Error} (RMSN) following the bi-level calibration with and without feedback function. For the first scenario (Figure \ref{fig:Speed HEATMAP}), the RSMN for the sequential framework is $0.298$, While for the proposed model, RMSN $= 0.134$. We also observe a noticeable improvement in using the feedback function for Scenario 2: $0.532$ and $0.191$ for the sequential and the proposed framework, respectively. 

\begin{figure}[!h]
    \centering
    \subfigure[Sequential calibration.]
    {
        \includegraphics[width=0.48\textwidth]{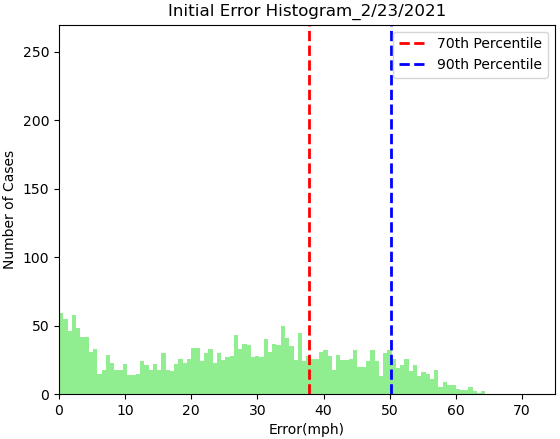}
        \label{fig:Speed_Error1}
    }
    \subfigure[Bi-level calibration.]
    {
        \includegraphics[width=0.48\textwidth]{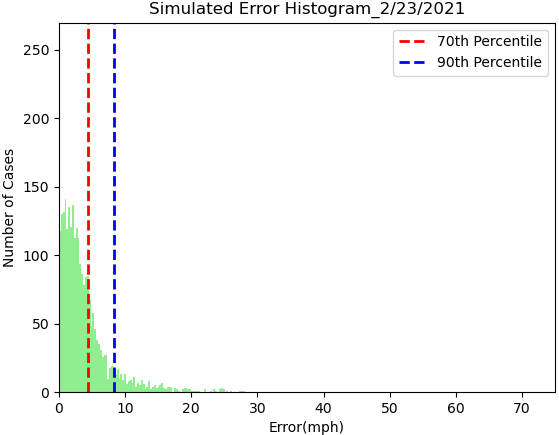}
        \label{fig:Speed_Error2}
    }
    \caption{Speed error histogram for the example of Stop-and-Go Waves scenario: The distribution of road segments based on their squared error values for speed, illustrating the precision of the optimal solution provided by our framework. Each bar in the histogram corresponds to the total count of segments (y-axis) with specific error values (x-axis), offering insights into the accuracy and reliability of the proposed model in complex traffic conditions.}
    \label{fig:Speed_Error}
\end{figure}

For evaluating the SPSA algorithms at the lower level, we assessed the solution quality across the stop-and-go waves scenario. Figure \ref{fig:Speed_Error} presents error histograms for speed. The x-axis displays the total speed deviation in miles per hour (mph), aggregated from all INRIX segments and across all time intervals. The y-axis shows the frequency of instances corresponding to each level of deviation, essentially representing the error distribution. 

Figure \ref{fig:Speed_Error1} illustrates the outcomes prior to applying the feedback mechanism, whereas Figure \ref{fig:Speed_Error2} presents the results post-feedback integration. Importantly, the red dash line marks the point where fewer errors occur in 70\% of instances, and the blue dash line indicates where 90\% of instances have reduced errors. One notable finding is the noticeable shift in the location of these dash lines, especially the one marking the 90\% threshold when the feedback function is activated during calibration. This emphasizes the critical role of the feedback mechanism in enhancing calibration accuracy, particularly in complex and highly congested scenarios. 

The importance of these results lies in their application to the background traffic simulation for the MVT field experiment. By carefully analyzing the observed congestion patterns, we have aimed to mitigate the addition of new congestion sources in this study, thereby maintaining a realistic and representative experimental setup. For example, across all scenarios, we consistently observed congestion occurring around Exit 59, which could be due to multiple factors such as lane changes, exits, and potentially other disruptions like accidents. Despite the limitations, these findings guided us in calibrating various scenarios to capture the background traffic, aiding in the design and pre-assessment of the MVT experiment.

\begin{figure}[H]
    \centering
    \includegraphics[width=0.48\textwidth]{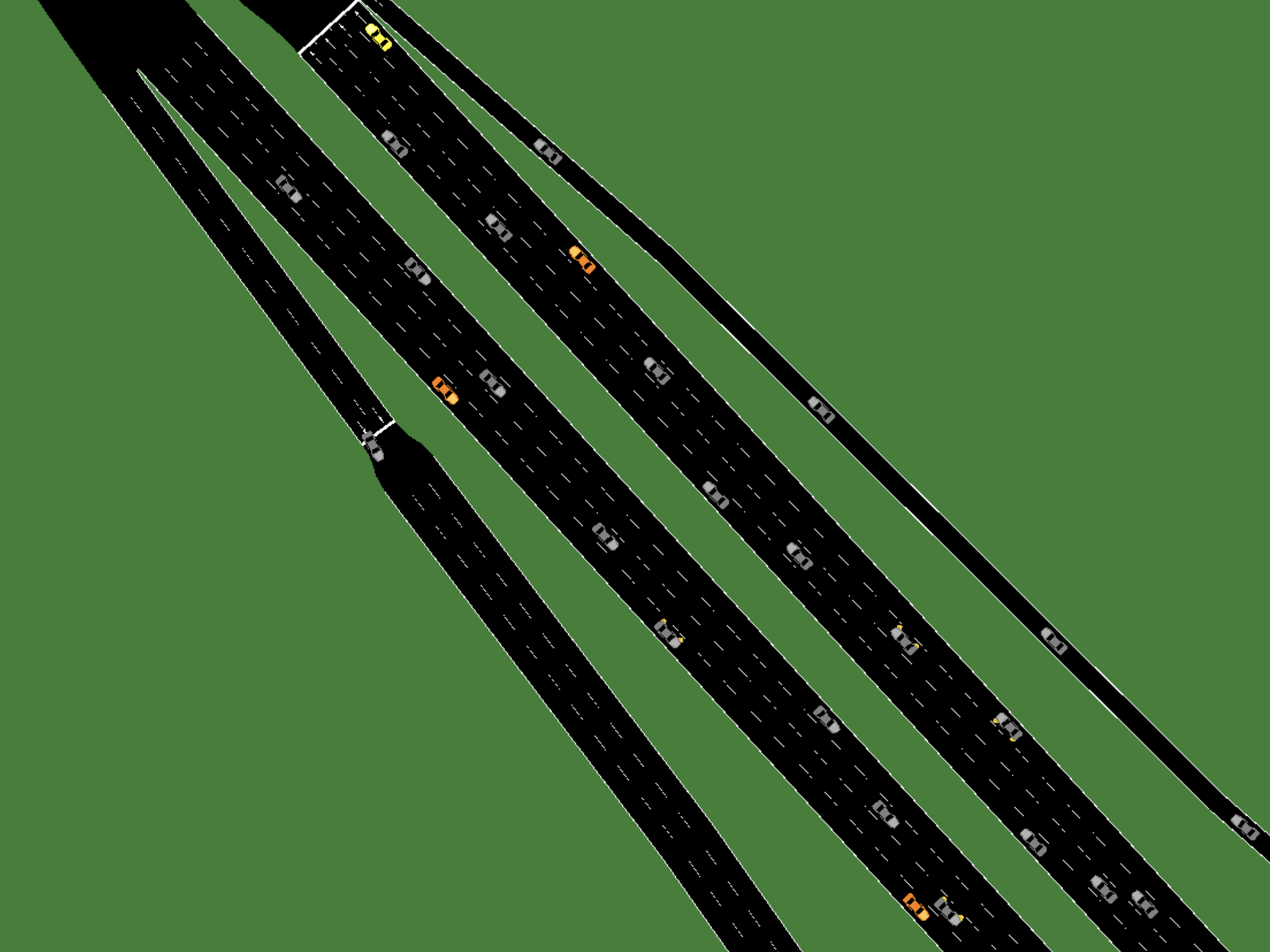}
    \caption{Snapshot of the I-24 simulation with AVs  (colored based on their route) and the background traffic (gray vehicles).}
    \label{fig:I24simwcars}
\end{figure}

This simulation tool is used to evaluate the impact of adding AVs. We test various routes and departure time distribution for AVs and investigate the potential overflow in different parts of the road network. Figure \ref{fig:I24simwcars} presents a snapshot of the microsimulation after adding AVs. The gray vehicles represent the background traffic, and AVs are colored based on their route. AVs are assigned to their planned lane and do not change except to use the off-ramps determined by their route. Note that we do not apply any restriction on the driving model and lane-changing behavior of background traffic. We calculate multiple macroscopic network measures (for example, network mean speed, density, and average travel time) and microscopic performance indicators (for example, average of space gaps, variance of space gaps, minimum of space gaps, and maximum of space gaps.) to compare scenarios with and without AVs in order to ensure that the final plan for adding all 100 AVs (even if they drive as HDVs) does not impact the traffic condition significantly. The final plan provides us with an estimation of the AV penetration rate during the MVT experiment.    

\begin{sidebar}{Macroscopic traffic models}
\noindent
by Dan Timsit, Mostafa Ameli, and Benedetto Piccoli

\sdbarinitial{T}\emph{he first macroscopic traffic model} was proposed
in the 1950s and known as the
Lighthill-Whitham-Richards (LWR) model \cite{lighthill1955,richards1956}:
\begin{equation}
\label{eq:LWR}
\begin{split}
\rho_t + (\rho v)_x &= 0 ,\\
\text{with}\quad v &= V(\rho) ,
\end{split}
\end{equation}
where $\rho$ is the car density, $v$ the average speed, and
the flux function $Q=Q(\rho)=\rho V(\rho)$ gives the number of vehicles passing by second and by meter.
The equation models the conservation of the number of cars and was inspired by fluid dynamics \cite{piccoli2006traffic}.\\
Data show that in low-density traffic there is a one-to-one correspondence between density and flux, which is lost in the congested regime. This leads to other models where the flux function depends on an additional variable $w$, such as
Aw–Rascle–Zhang model \cite{AwRascle00,ZHANG2002,yu2019traffic, yu2019boundary, belletti2015prediction}, Phase-Transition models \cite{Blandin2011,Colombo2003},
and General Second Order Models \cite{fan_collapsed_2017,FanHertySeibold14,lebacque2007generic}.  %yu2019traffic yu2019boundary fan2017collapsed 
%yu2019traffic,belletti2015prediction,yu2019boundary,fan2017collapsed,
These models take the form of a system of two Partial Differential Equations (PDEs):
\begin{equation}
\label{eq:generic2}
\begin{split}
\rho_t + (\rho v)_x &= 0 ,\\
w_t + v w_x & = 0 ,\\
\text{with}\quad v &= V(\rho,w) .
\end{split}
\end{equation} 
%where $w$ is a new variable describing (and not explaining) the flux $Q$ which is now given by $Q=Q(\rho, w)=\rho V(\rho, w)$.
%For numerical simulations, we prefer to use another variable: 
One can write the system \eqref{eq:generic2}
in conservative form using the variable $y=\rho \cdot w$:
\begin{equation}
\label{eq:ARZ_with_y}
\begin{split}
\rho_t + (\rho v)_x &= 0 ,\\
y_t + (v y)_x & = 0 ,\\
\text{with}\quad v &= V(\rho,\frac{y}{\rho}) .
\end{split}
\end{equation}
The main modeling ingredient is the velocity function $V$,
which for the ARZ model was chosen as:
\begin{equation}
\label{eq:velocity_arz}
V(\rho,w) = V_\text{eq}(\rho)+\left(w-V_\text{eq}(0)\right),\quad \text{for $\rho\in [0,\rho_\text{max}]$.} 
\end{equation}
The GSOM model uses a general $V$ with the assumptions
that $w \rightarrow Q(\rho, w)$ is injective, thus the relation between the density and the flux determines uniquely $w$.
The model proposed in \cite{fan_collapsed_2017},
called collapsed Generalized ARZ, discards this last hypothesis and assumes that in free-flow regime ($\rho <\rho_c$, for some $\rho_c>0$) the fundamental diagram does not depend on $w$.
%, but that in congested areas $\rho>\rho_c$ this dependence reappears.
Thus the fundamental diagram takes the form:
\begin{equation}
\label{eq:cgarz}
\left\{\begin{array}{rl}
\rho_t + (\rho v)_x &= 0 ,\\
y_t + (y v)_x & = 0 ,\\
v=V(\rho,y/\rho)&=\left\{\begin{array}{rl}
	        V_{\text{f}}(\rho) , &~\text{ if}~ 0\le\rho\le\rho_{\text{f}} ,\\
	         V_{\text{c}}(\rho,y/\rho) , &~\text{ if}~ \rho_{\text{f}}<\rho\le \rho_{\text{max}} .\end{array} \right.
\end{array} \right.	
\end{equation}
In each case, one assumes the families of fluxes to satisfy $Q(0, w)=Q(\rho_{\mbox{max}}, w)=0$. This corresponds to either no vehicles or vehicles not moving at maximal density.\\
%Here are some particular families of fundamental diagrams which are adapted to the aforementioned models:
%\begin{itemize}
%    \item CGARZ:  \begin{equation}
% \label{eq:congesteddiagram0}
% \begin{split}
%   Q_{\text{c}}(\rho,w) &= Q_{\sigma(w),\mu(w)}\left(\rho\right)\\
%    &=-\sigma(w) c \left(\left(\tfrac{\rho-\mu(w)}{\sigma(w)}\right)\tan^{-1}\left(\tfrac{\rho-\mu(w)}{\sigma(w)}\right) - \right. \\
%    &\qquad \left. \tfrac{1}{2}\ln\left(1+\left(\tfrac{\rho-\mu(w)}{\sigma(w)}\right)^2\right)\right) \\
%    &\qquad + b \rho + c ,
%\end{split}
%\end{equation}
%where $b$, $v_f$, $I$ and $c$ depend on the other (free) parameters.
%\item GARZ: \begin{equation}
% \label{eq:garzfd}
% \begin{split}
%   Q(\rho,w)
%   &=Q_{\alpha(w),\lambda(w),p(w)}(\rho)\\
%   &=\alpha\left(w\right)\left(a_w+(b_w-a_w)\tfrac{\rho}{\rho_{max}}-\sqrt{1+y_w^2}\right),
% \end{split}
%\end{equation}
%\end{itemize}
%    
%\begin{itemize}
%\item ARZ: $V(\rho, w)$ is an affine function of $w$
%
%\begin{equation*}
%\label{eq:arz_eq_v}
%\begin{split}
%V_{\text{eq}}\left(\rho\right)=\frac{Q\left(\rho, w_{\text{eq}}\right)}{\rho},
%\end{split}
%\end{equation*}
%\item LWR: $Q(\rho) = Q(\rho, w_{eq})$
%\end{itemize}
Solutions to conservation laws may exhibit traveling discontinuities in finite time (even for smooth initial data), called shocks. The latter correspond to queues in real traffic, due to traffic lights or network features. 
%Explicit solutions to these models for unconstrained initial datum may not exist if the flux form is not accommodating. We turn to a specific type of initial datum which casts our initial LWR as a Riemann problem:
Solutions to general Cauchy problems are constructed based on solutions to the so-called Riemann problems, which for the LWR model read:
\begin{align}\label{eq:generic}
    \rho_t + (\rho v)_x &= 0 ,\\
    \text{with}\quad v &= V(\rho,w) ,\\
    \rho(x, 0) &= 
    \begin{cases}
        \rho_l & \text{ for } x<0, \\
        \rho_r & \text{ for } x>0.
    \end{cases}
\end{align}
This can be solved explicitly with rarefaction waves if $\rho_l<\rho_r$ and with shock waves if $\rho_l>\rho_r$.
The case of systems is move involved and we refer the reader to \cite{bressan2000hyperbolic} and references therein.\\
To deal with the case of road networks, the theory of conservation laws has been extended to graphs, see \cite{Coclite2005}. For a well-defined theory, one usually supplies a system of conservation laws with dynamics rules at junctions:\\
(A) The traffic flow is distributed from incoming roads to the outgoing roads linearly according to a matrix $A$.\\
(B) The flow is maximized while respecting rule (A).\\
Rules (A) and (B) define a linear programming problem at each junction, thus allowing for efficient solutions and simulations on road networks. Other researchers proposed more complicated rules. The main idea is to identify physically-motivated rules that allow to solve uniquely Riemann problems at junctions, which are defined by initial data constant on each road. Then solutions can be defined on the whole network by combining solutions to Riemann problems at junctions and along the roads. We refer the reader to \cite{garavello2016models} for a comprehensive discussion. 
\\
% A wave-front tracking algorithm can be developed to solve the original PDE problem. The continuous (or not) initial datum is approached by a piece-wise constant function.
% \\Unfortunately this method only applies to LWR, we can however discretize time and 
% \begin{equation}
% \label{eq:sd_2}
%  \begin{split}
%  \rho^{n+1}_j &= \rho^{n}_j + \frac{\Delta t}{\Delta x} \left(F^{\rho,n}_{j-1/2} - F^{\rho,n}_{j+1/2}\right) ,\\
%  y^{n+1}_j &= y^{n}_j + \frac{\Delta t}{\Delta x} \left(w^n_{j-1} F^{\rho,n}_{j-1/2} -w^n_{j} F^{\rho,n}_{j+1/2}\right) .
%  \end{split}
% \end{equation}
% where $F^{\rho, n}_{j-1/2} = \min(S(\rho^n_{j-1}), R(\rho_j^n))$ and $F^{\rho, n}_{j+1/2} = \min(S(\rho^n_{j}), R(\rho^n_{j+1}))$
% and where S and R are the so-called sending and receiving functions which are defined by the flux function $Q$.
While macroscopic models have been used to establish large-scale traffic features, such as average speed and density, they cannot be used to investigate micro-scale features, such as the number of vehicles in a queue. There are multiple studies in the literature to discretize the macroscopic models based on statistical methods (see, for example, \cite{work2008ensemble, work2010distributed, seo2016filter}). However, most of the microscopic features, such as the interaction of vehicles and detailed intersection representation, cannot be captured with such methodologies.
On the other side, macroscopic models are particularly useful for data fitting \cite{piccoli2015second}, representing waves \cite{Flynn09}, and solving optimization problems \cite{cascone2008circulation,Rarita11}. 
\end{sidebar}

% \begin{figure}
%     \centering
%     \subfigure[First caption]
%     {
%         \includegraphics[width=1.0in]{figures/RoutesWithQRCode_UpdatedCarFormation5.png}
%         \label{fig:first_sub}
%     }
%     \\
%     \subfigure[Second caption]
%     {
%         \includegraphics[width=1.0in]{figures/RoutesWithQRCode_UpdatedCarFormation5.png}
%         \label{fig:second_sub}
%     }
%     \subfigure[Third caption]
%     {
%         \includegraphics[width=1.0in]{figures/RoutesWithQRCode_UpdatedCarFormation5.png}
%         \label{fig:third_sub}
%     }
%     \caption{Common figure caption.}
%     \label{fig:sample_subfigures}
% \end{figure}
% %% subfigure example from Samaei et. al. TRB
% \begin{figure}[!h]
%  \centering
%  \begin{subfigure}[b]{0.42\textwidth}
%    \includegraphics[width=\textwidth]{Figures/map_network.png}
%    \caption{Mapping data \copyright Google maps}
%    \label{fig:flow-error}
%  \end{subfigure}
%  \begin{subfigure}[b]{0.57\textwidth}
%    \includegraphics[width=\textwidth]{Figures/network_SUMO.png}
%    \caption{SUMO network}
%    \label{fig:network}
%  \end{subfigure}
%  \caption{I-24 highway}
%  \label{network}
%  \end{figure}
%  %% end of subfigure example from TRB
\begin{sidebar}{Driver safety procedures and training}
\noindent
by Sean McQuade and Riley Wagner

\setcounter{sfigure}{7}
\renewcommand{\thesfigure}{S\arabic{sfigure}}
\noindent
\sdbarfig{\includegraphics[width=19.0pc]{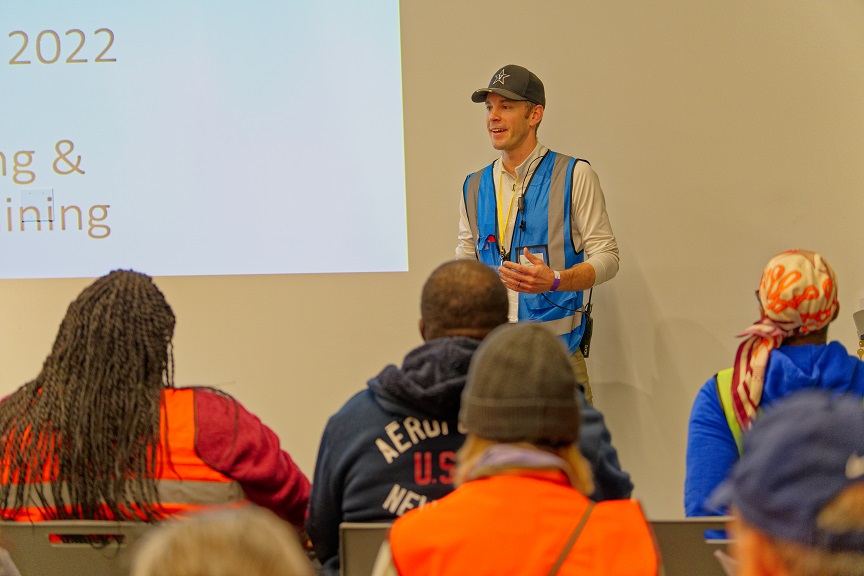}}{A training session with orange route drivers. The session is presented by Professor Dan Work. \label{fig:DriverTraining}}

\sdbarinitial{T}\emph{he AV drivers} were trained to activate the Adaptive Cruise Control (ACC) system in their vehicles after they safely entered their assigned lane. They were taught their routes, between which exits they should repeatedly loop, and which exit to use to return to \textit{Field Headquarters} (FHQ). Professor Work came up with the simple acronym SNAP, which stands for Safety (a successful test is one where no one gets injured), Navigation (drivers should drive on their assigned routes), ACC (drivers should activate cruise control when safe to do so on the Westbound direction), Position (drivers should remain in the assigned lane). The order of SNAP indicates the priority of the four instructions.

Driver breaks were necessary so that drivers would only drive during part of the experiment and then rest while another driver substituted for them. This procedure ensured that drivers were alert and safe during the entire experiment. A specific safe traffic flow through the FHQ parking lot was designed to account for vehicles returning to the lot and pedestrians walking to/from AVs. 

\vspace{0.2cm}

\noindent
\sdbarfig{\includegraphics[width=19.0pc]{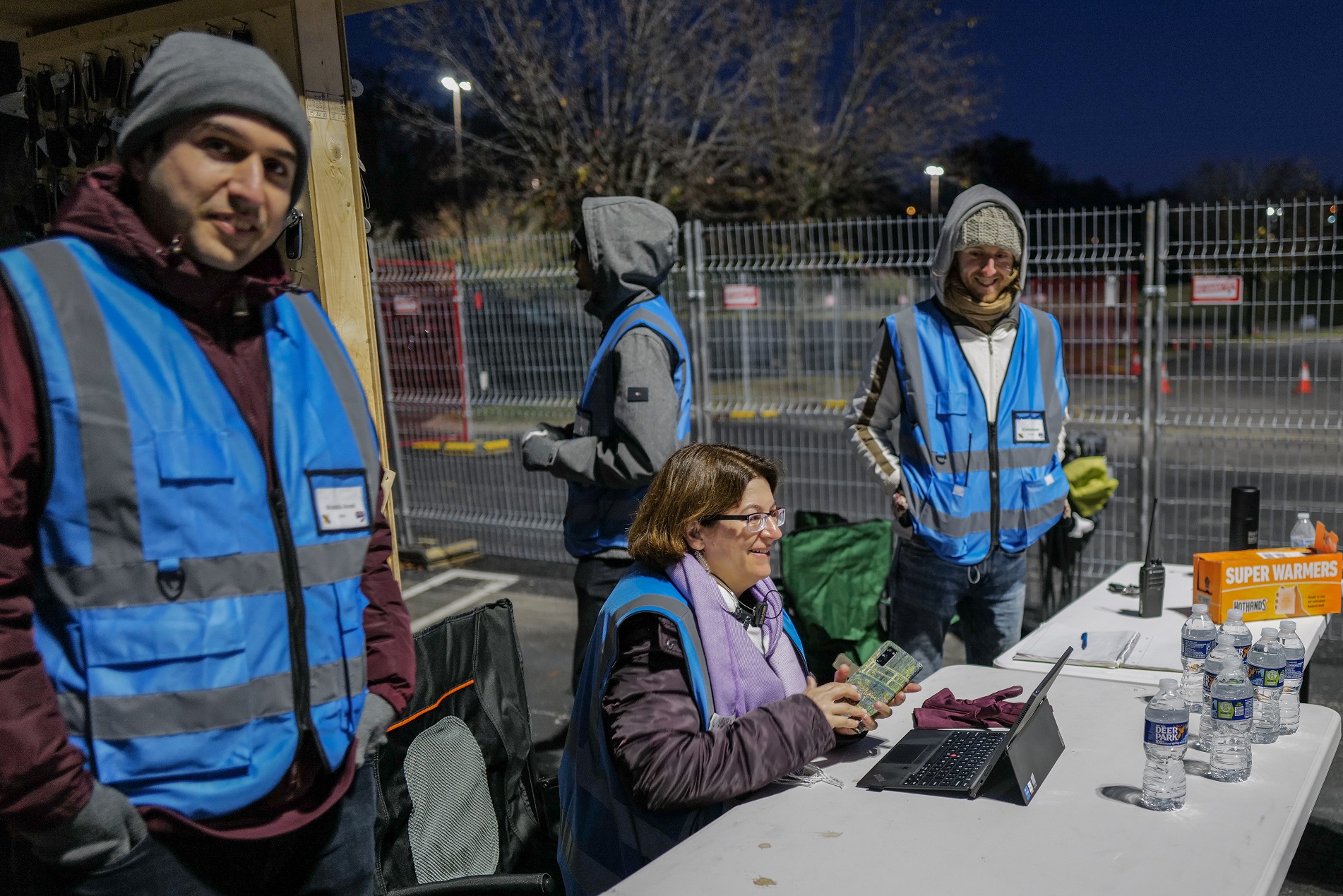}}{The key distribution table preparing to hand out keys and record which drivers received them. Mostafa Ameli (left) Sharon Hornstein (center) and George Gunter (right) are prepared to manage hundreds of key distributions as vehicles were rotating in and out of the parking lot to switch drivers.\label{fig:KeyDesk}}

The drivers were trained to follow the instructions of designated persons guiding vehicles through the parking lot. This was necessary to maximize the safety of drivers walking to vehicles and parking lot crew.
% \sdbarfig{\includegraphics[width=19.0pc,trim={1cm 0cm 2.6cm 0}, clip]{figures/ParkingLotFlow.PNG}}
% \sdbarfig{\includegraphics[width=19.0pc,trim={3cm 6cm 1cm 4cm 0}, clip]{figures/Lot Photo.jpg}}
% {Flow of AVs through our headquarters parking lot. Top: the planned positions of vehicles and crew designed to allow drivers of the Orange route and Yellow route to return safely and substitute for each other. Bottom: a photo of the lot with Orange route vehicles on the left, Yellow on the right.}

\sdbarfig{\includegraphics[width=19.0pc]{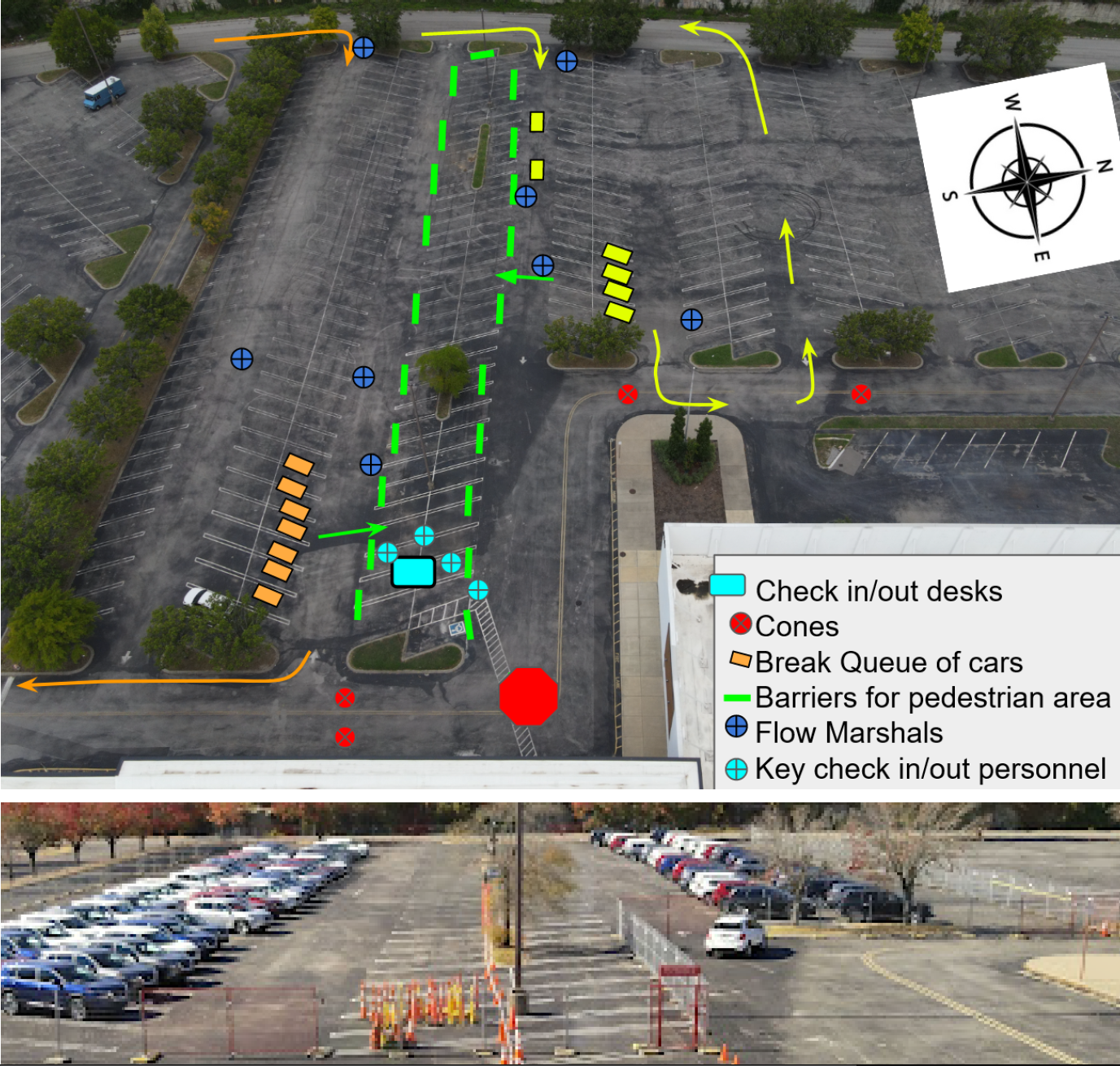}}{Flow of AVs through the FHQ parking lot. Top: the planned positions of vehicles and crew designed to allow drivers of the orange route and yellow route to return safely and substitute for each other. Bottom: a photo of the lot with orange route vehicles on the left and yellow on the right.}

Once a driver returned from the parking lot and exited their vehicle, they turned in their vehicle key to a desk set up in the lot, then headed to the break area inside the FHQ to rest. Another available driver would be invited to the downstairs lobby in anticipation of the returning vehicles. They would be instructed to come to the lot in anticipation of a vehicle returning. Communication between the parking lot team, the lobby attendant, and upstairs driver attendants using ham radios helped to minimize the amount of time a vehicle was waiting for a replacement driver.\\

% \noindent
% \sdbarfig{\includegraphics[width=19.0pc]{figures/CarsInLotCropped}}{The key distribution table preparing to hand out keys and record which drivers received them. Front: Mostafa Ameli has organized the keys on the peg board based on the cars' position in the lot.\label{fig:CarsInLotCropped}}

Driver training culminated in a test drive lap of the particular route that the driver trained for to familiarize themselves with the vehicles, routes, and exits. If a driver had any problems during training or the test, they were instructed to exit the highway to a safe location and call a phone number if they could not return safely to FHQ. We set up this number to receive several calls simultaneously should the need arise.
\end{sidebar}

\section{Penetration rate estimations}\label{sec:PenetrationRate}

% The background traffic (vehicles that are not AVs) during AM peak congestion was recorded for a previous year in the TDOT report: ``I-24 Ramp Metering Study (From I-840 to I-40) Rutherford and Davidson Counties
% Traffic Operations Report
% JULY 2021.'' We show data from this report in Figure \ref{fig:TDOT_I24_volume}, which gives counts of vehicles passing through roads during the times of 6:30am - 7:30am and 7:30am - 8:30am. These two hours are considered ``AM peak traffic.''  

% \FloatBarrier
Two routes (orange and yellow presented in the MVT testbed location") are determined based on the evaluation of the background traffic. To find the proportion of AVs to bulk traffic that results from driving the chosen routes during AM peak traffic (heading westbound), we calculated how many times a control vehicle was expected to travel its entire route in one hour based on the simulation results. If this was 42 minutes, as we initially estimated for the orange route, then we derive that a control vehicle to make 1.43 westbound drives per hour. We estimate that a control vehicle doing a 42 minute loop of consecutive drives would contribute 1.43 vehicles westbound per hour. We termed this hourly contribution, ``effective vehicles per hour.'' The background traffic was scaled by $\frac{3}{4}$ since we were only driving in lanes 2, 3, and 4 (not the far left HOV lane). Lanes are numbered according to TDOT; they are shown in Figure \ref{fig:Routes} in the sidebar ``MegaVanderTest testbed location.''

Using a theoretical release schedule (Figure \ref{fig:Gannt}) allowed us to consider how many reserve drivers we would need for each route. This schedule assumes the AVs exit the \textit{Field Headquarters} (FHQ) at regular intervals, alternating between the orange route and the yellow route. The optimized simulation allowed us to compute expected times for a complete loop and leading to this ideal schedule. All drivers keep perfect ordering from when they leave to when they return. This was used to estimate the number of extra drivers needed to keep all vehicles on the road during breaks. Note that after a break, the drivers who return from the break get into an AV with the same color route, but not the same AV, and not necessarily the same lane assignment. For this reason, we indicated the lane inside each AV. Drivers going on their break sufficiently late in the order would not necessarily get back into an AV but would remain in the FHQ in case an AV of their color route returned.

\setcounter{figure}{10}
\begin{figure}[H]
    \centering
    \includegraphics[width=0.48\textwidth]{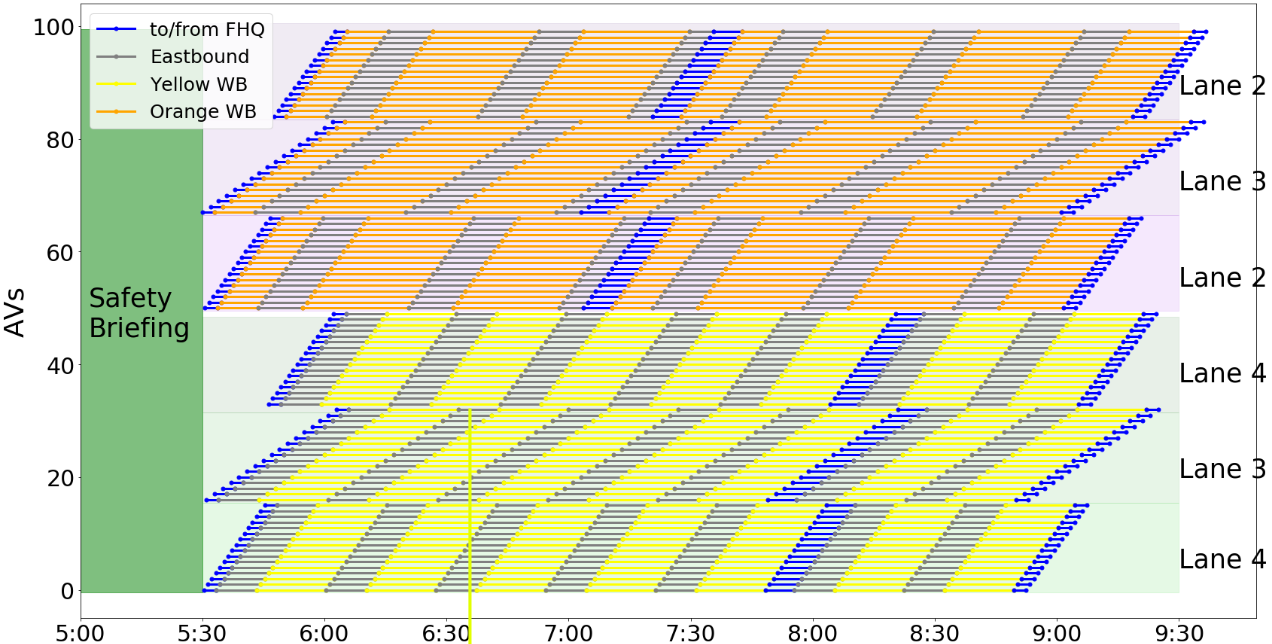}
    \caption{An planned schedule of the first 100 drivers and their breaks. The lane assignments on the right show the lanes of the first AVs driven by the driver. The orange/yellow segments indicate orange/yellow route AVs driving Westbound, the gray segments indicate them driving eastbound back to start another loop, and the blue segments indicate drivers returning to the lot for a break.}
    \label{fig:Gannt}
\end{figure}

% \begin{figure}
%     \scalebox{0.57}{
%     \begin{tikzpicture}[node distance=2cm, text width=2.5cm, align=center]
%     \tiny
%     % Nodes
%     \node (road) at (-1,-1.5) {\includegraphics[width=14cm, trim = {2cm 0 0 0}] {figures/PenetrationRate_blank1.png}};
    
%     \node[text width=20mm] (Haywood) at (-1.4,-.4 ) {\Large Haywood};
%     \node (Bell) at (1.3,-.4 ) {\Large Bell Rd};
%     \node[text width=20mm] (Hickory) at (5.7,-.3 ) {\Large Hickory Hollow};
%     \node[text width=20mm] (OldHickory) at (8.9,-.3 ) {\Large Old Hickory};
%     \node (stats) at (-.4,-1.2 ) {\large $1.25\%|1.32\%$};
%     \node (stats) at (3.4,-1.2 ) {\large $1.34\%|1.36\%$};
%     \node (stats) at (7.0,-1.2 ) {\large $2.79\%|2.78\%$};
%     \node (stats) at (10.3,-1.2 ) {\large $2.99\%|2.78\%$};
    
%     \node (route) at (-0.5,-2.7 ) {\large Orange Route Only};
%     \node (route) at (3.5,-2.7 ) { \large Orange Route Only};
%     \node (route) at (7.2,-2.7 ) {\large Orange and Yellow};
%     \node (route) at (10.4,-2.7 ) {\large Orange and Yellow};
%     %\node (location) at (-3,-3) {Experiment Location};
    
%     \end{tikzpicture}
%     }
%     \end{figure}

\begin{figure}[H] 
    \centering
    \includegraphics[width=0.48\textwidth, trim={0.2cm 0cm 0.02cm 0cm}, clip]{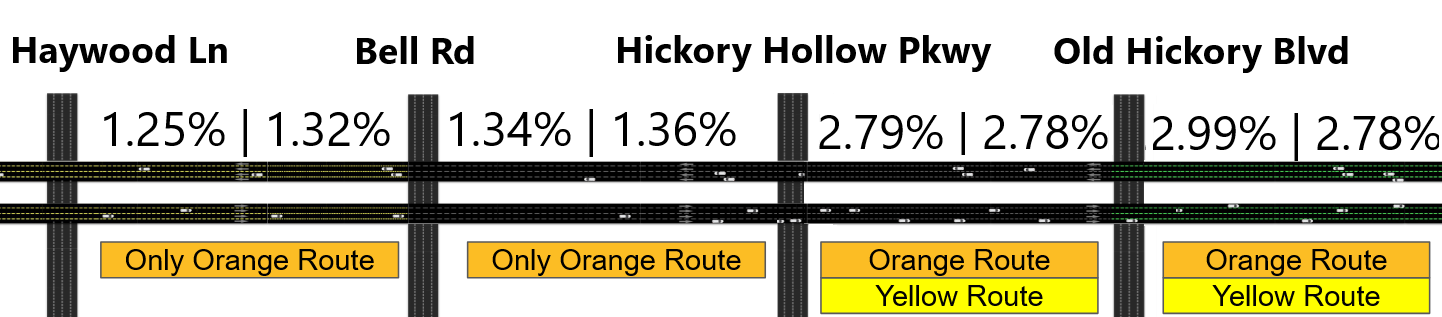}
    \caption{In preparation for the experiment, estimations were made on sections of the highway with camera coverage to make sure there are a sufficient number of control vehicles compared to the background traffic conditions. We calculate the proportion during the two peak hours 6:30am - 8:30am. The left (right) number shows the estimation for the first (second) hour. }
    \label{fig:PenetrationRate}
\end{figure}

The penetration rates described ``temporal'' estimates as opposed to penetration based on a given road segment. When a spatial estimate through various simulation scenarios for penetration rate was more suitable, we estimated the number of other vehicles that would be between two AVs assuming uniformly spaced AVs.

% \section{Impact of adding AVs on the traffic variables}

% This section will conclude and validate the design of the experiment with a table includes the traffic monitoring KPIs for two scenarios: Background traffic and MVT simulation scenario (adding 100 AVs with their actual routes and departure time).

%%%%%%%%%%%%%%%%%%%%%%%%%%%%%%%%%%%%%%%%%%%%%%%
%                7.Conclusion                 %
%%%%%%%%%%%%%%%%%%%%%%%%%%%%%%%%%%%%%%%%%%%%%%%
\section{Conclusion}
This article is a tutorial encompassing the intricacies of planning, designing, and executing large-scale field experiments in live traffic control, focusing on the MegaVanderTest conducted by the CIRCLES Consortium. Stemming from preliminary research highlighting the potential of AVs to mitigate recursive waves in simpler settings, MVT presents a complex achievement by applying these insights to real-world, high-density traffic networks. The MVT experiment, as the largest live traffic control experiment at the time, was conducted southeast of Nashville, TN, USA, and involved 100 vehicles executing various control algorithms to alleviate stop-and-go traffic waves.

The article delves into the comprehensive three-year planning process leading up to MVT. This rigorous preparation included choosing optimal dates and times for the experiment, meticulous route planning, hardware installations, as well as the development and simulation of control algorithms. These elements were planned to ensure the effective deployment and assessment of AVs within a bustling traffic environment. A cornerstone of this work is the innovative bi-level calibration framework with an integrated feedback function. This method sets a new benchmark by effectively handling disparate data sets with varied spatiotemporal characteristics, thereby allowing for a more accurate simulation of real-world traffic conditions to evaluate the impact of control vehicles. We validated this calibration approach using data from a six-mile stretch of Nashville's I-24 highway, underscoring its efficacy in bridging the gap between simulated and real-world traffic data. Such precision in calibration proved crucial for the successful design and execution of the MVT experiment.

Although these findings are promising, the article acknowledges existing limitations, particularly in the realm of input data quality. The uncertainties around the underlying reasons for specific congestion patterns remain a challenge. Future work aims to enhance the calibration model by incorporating higher-quality, lane-specific data from I-24 MOTION~\cite{gloudemans2020interstate,gloudemans202324}. Such data will offer a more comprehensive understanding of traffic conditions, thus paving the way for even more accurate and realistic simulations.

 % provides a detailed roadmap for similar large-scale experiments, introduces an innovative calibration technique adaptable to real-world complexities, and sets the stage for future advancements in intelligent and sustainable traffic management. As we continue to refine our methodologies and integrate more robust data, we anticipate that this work will act as a foundational resource in the ongoing quest for more efficient and sustainable transportation systems.

\section{Acknowledgement}

This material is based upon work supported by the National Science Foundation under Grants CNS-1837244 (A. Bayen), CNS-1837652 (D. Work), CNS-1837481 (B. Piccoli), CNS-1837210 (G. Pappas), CNS-1446715 (B. Piccoli), CNS-1446690 (B. Seibold), CNS-1446435 (J. Sprinkle, R. Lysecky), CNS-1446702 (D. Work), CNS-2135579 (D. Work, A. Bayen, J. Sprinkle, J. Lee). This material is based upon work supported by the U.S.\ Department of Energy’s Office of Energy Efficiency and Renewable Energy (EERE) under the Vehicle Technologies Office award number CID DE--EE0008872. The views expressed herein do not necessarily represent the views of the U.S.\ Department of Energy or the United States Government. This material is based upon work supported by the Vice-President International of Université Gustave Eiffel, France, under "OII1 - Initiative ciblée et ponctuelle" funding – n° 3074.

\section{Authors}

\begin{IEEEbiography}{{M}ostafa Ameli}{\,}(mostafa.ameli@univ-eiffel.fr) is an Assistant Professor in applied mathematics, computer science, and transportation science at the Transportation Engineering and Computer Science Lab (GRETTIA), University Gustave Eiffel Paris, France. With a keen interest in the crossroads of operations research and computer science, he focuses on deploying mathematical, machine-learning, and computational tools to design smarter, more resilient transportation systems. Since 2022, he is also a Research Affiliate with the Department of Electrical Engineering and Computer Sciences at the University of California, Berkeley. 
\end{IEEEbiography}

% \begin{IEEEbiography}{{S}ean McQuade}{\,}(sean.mcquade@rutgers.edu) 
% Sean McQuade received a B.S.~degree in Mathematics from Virginia Polytechnic Institute and State University. He received a Ph.D.~degree in Computational and Integrative Biology from Rutgers–Camden. He currently works as a Postdoc at Rutgers–Camden as a senior researcher.
% \end{IEEEbiography}\\

\begin{IEEEbiography}{{J}onathan W.~Lee}{\,}(jonny5@berkeley.edu) received the B.S.~degree in engineering physics from the University of California, Berkeley and M.S.~and Ph.D.~degrees in mechanical engineering from Rice University. At Sandia National Laboratories (2011--13), he completed his postdoctoral appointment studying electrical and material properties via molecular dynamics simulations. He subsequently served as a senior data scientist and product manager on various teams at Uber Technologies, Inc.~(2014-2019). Since 2019, he has served as an engineering manager at the University of California, Berkeley and the program manager and Chief Engineer of CIRCLES.
\end{IEEEbiography}

\begin{IEEEbiography}{{M}aria Laura Delle Monache}{\,}(mldellemonache@berkeley.edu) is an assistant professor in the Department of Civil and Environmental Engineering at the University of California, Berkeley. Dr. Delle Monache’s research lies at the intersection of transportation engineering, mathematics, and control and focuses on modeling and control of mixed autonomy large-scale traffic systems.
\end{IEEEbiography}

\begin{IEEEbiography}{Benjamin Seibold}{\,}(seibold@temple.edu) 
is a Professor of Mathematics and Physics, and the Director of the Center for Computational Mathematics and Modeling, at Temple University. His research areas, funded by NSF, DOE, DAC, USACE, USDA, and PDA, are computational mathematics (high-order methods for differential equations, CFD, molecular dynamics) and applied mathematics and modeling (traffic flow, invasive species, many-agent systems, radiative transfer).
\end{IEEEbiography}

\begin{IEEEbiography}{{D}an Work}{\,}(dan.work@vanderbilt.edu) is a Professor of Civil and Environmental Engineering and the Institute for Software Integrated Systems, at Vanderbilt University. His interests are in transportation cyber-physical systems.
\end{IEEEbiography}

\begin{IEEEbiography}{{J}onathan Sprinkle}{\,}() 
Jonathan Sprinkle is a Professor of Computer Science at Vanderbilt University since 2021. Prior to joining Vanderbilt he was the Litton Industries John M. Leonis Distinguished Associate Professor of Electrical and Computer Engineering at the University of Arizona, and the Interim Director of the Transportation Research Institute. From 2017-2019 he served as a Program Director in Cyber-Physical Systems and Smart \& Connected Communities at the National Science Foundation in the CISE Directorate. 
\end{IEEEbiography}

\begin{IEEEbiography}{{B}enedetto Piccoli}{\,}(piccoli@camden.rutgers.edu) is University Professor at Rutgers University-Camden. He also served as Vice Chancellor for Research. He received his Ph.D. degree in applied mathematics from the Scuola Internazionale Superiore di Studi Avanzati (SISSA), Trieste, Italy, in 1994. He was a Researcher with SISSA from 1994 to 1998, an Associate Professor with the University of Salerno from 1998 to 2001, and a Research Director with the Istituto per le Applicazioni del Calcolo “Mauro Picone” of the Italian Consiglio Nazionale delle Ricerche (IAC-CNR), Rome, Italy, from 2001 to 2009. Since 2009, he has been the Joseph and Loretta Lopez Chair Professor of Mathematics with the Department of Mathematical Sciences, Rutgers University–Camden, Camden, NJ, USA.
\end{IEEEbiography}

\begin{IEEEbiography}{{A}lexandre M. Bayen}{\,}(bayen@berkeley.edu)
is the Associate Provost for Moffett Field Program Development at UC Berkeley, and the Liao-Cho Professor of Engineering at UC Berkeley. He is a Professor of Electrical Engineering and Computer Science, and of Civil and Environmental Engineering (courtesy). He is a Visiting Professor at Google. He is also a Faculty Scientist in Mechanical Engineering, at the Lawrence Berkeley National Laboratory (LBNL). From 2014 - 2021, he served as the Director of the Institute of Transportation Studies at UC Berkeley (ITS). 
\end{IEEEbiography}

% \end{thebibliography}

\nocite{*}
\bibliographystyle{IEEEtran}
\bibliography{references}

\endarticle

\end{document}